\PassOptionsToPackage{table}{xcolor}
\documentclass{iucrjournals}

\usepackage{amssymb}
\usepackage{amsmath}
\usepackage{booktabs}
\usepackage{boldline}
\usepackage{colortbl}
\usepackage{siunitx}
\usepackage{placeins}
\usepackage[table]{xcolor}
\newcommand{\PSIfour}{\mbox{P{\scalebox{.82}{SI}}4}}

\title{Toward Routine CSP of Pharmaceuticals: A Fully Automated Protocol Using Neural Network Potentials}

\author[a]{Zachary L. Glick\IUCrCemaillink{zach@lavo.ai}\IUCrOrcidlink{0000-0003-0900-2849}}%
\author[a]{Derek P. Metcalf\IUCrCemaillink{derek@lavo.ai}\IUCrOrcidlink{0000-0003-2395-3229}}%
\author[a]{Scott F. Swarthout\IUCrCemaillink{scott@lavo.ai}}%

\affil[a]{Lavo Life Sciences, USA}

\begin{document} 
\maketitle 

\begin{synopsis}
A fully automated, low-cost crystal structure prediction (CSP) protocol, powered by a purpose-built neural network potential, is developed and demonstrated to rapidly and accurately identify polymorphs of pharmaceutical compounds.
This protocol is validated with the largest pharmaceutical CSP benchmark to date, paving the way for the routine use of computational solid-form screening in drug development.
\end{synopsis}

\begin{abstract}
Crystal structure prediction (CSP) is a useful tool in pharmaceutical development for identifying and assessing risks associated with polymorphism, yet widespread adoption has been hindered by high computational costs and the need for both manual specification and expert knowledge to achieve useful results.
Here, we introduce a fully automated, high-throughput CSP protocol designed to overcome these barriers. 
The protocol's efficiency is driven by Lavo-NN, a novel neural network potential (NNP) architected and trained specifically for pharmaceutical crystal structure generation and ranking.
This NNP-driven crystal generation phase is integrated into a scalable cloud-based workflow. 
We validate this CSP protocol on an extensive retrospective benchmark of 49 unique molecules, almost all of which are drug-like, successfully generating structures that match all 110 $Z' = 1$ experimental polymorphs. 
The average CSP in this benchmark is performed with approximately 8.4k CPU hours, which is a significant reduction compared to other protocols.
The practical utility of the protocol is further demonstrated through case studies that resolve ambiguities in experimental data and a semi-blinded challenge that successfully identifies and ranks polymorphs of three modern drugs from powder X-ray diffraction patterns alone. 
By significantly reducing the required time and cost, the protocol enables CSP to be routinely deployed earlier in the drug discovery pipeline, such as during lead optimization. Rapid turnaround times and high throughput also enable CSP that can be run in parallel with experimental screening, providing chemists with real-time insights to guide their work in the lab.
\end{abstract}


\section{Introduction}

Approximately 50\% of drugs are known to exhibit polymorphism, a phenomenon in which the same drug molecule can be observed in multiple geometrically distinct crystalline arrangements, referred to as polymorphs or polymorphic forms. \cite{C5CS00227C}
The ubiquity of polymorphism is hugely consequential for drug developers.
Because different polymorphs of the same drug may exhibit different physical properties (solubility, hygroscopicity, stability, compressibility, etc.), polymorphism presents an additional tunable variable in the formulation of the drug. \cite{Pudipeddi2005,Listiohadi2008}
For example, famotidine (brand name Pepcid) is formulated using metastable Form B over the slightly more stable Form A due to the faster dissolution rate and better manufacturability of Form B. \cite{Lin2014, EGART2014347}
More often than not, however, when formulating a drug, the possibility of polymorphism is a liability rather than a benefit.
In the infamous case of ritonavir, the thermodynamically more stable Form II unexpectedly appeared well after the drug reached the market. The much lower solubility and slower dissolution of this late appearing Form II compared to the originally marketed Form I caused a dramatic loss of oral bioavailability, forcing Abbott to withdraw and subsequently reformulate Norvir capsules. \cite{Bauer2001}
In addition to formulation concerns, there are regulatory requirements and intellectual property risks that arise from polymorphism. 
The FDA requires developers to identify and characterize all relevant polymorphic forms of the drug substance and to justify the selection of the specific form used in the product. \cite{Elder2017,USFDA2010}
Because individual polymorphic forms can be patented separately and can extend market exclusivity, they often become focal points for litigation. \cite{kattan2015what}

Given the critical importance of polymorphism, drug developers are strongly incentivized to rapidly and comprehensively map out every solid form of a candidate compound using systematic polymorph screening. 
In these screens, the drug is subjected to a broad array of solid-form generation techniques to identify all accessible crystalline forms. Such techniques include crystallization under various solvents, temperatures, pressures, and supersaturation levels; mechanochemical treatments such as grinding and milling; thermal routes including melt quenching and thermal cycling; seeding and vapor-phase methods; even micro- and hyper-gravity experiments. \cite{Neumann2008, Hilfiker2006PolymorphismIndustry, C5CS00227C, orbital-ritonavir}
Analytical methods such as powder X-ray diffraction (PXRD), single-crystal X-ray diffraction (SCXRD), differential scanning calorimetry (DSC), and solid-state NMR are then used to characterize each polymorph, while slurry and stress-testing experiments establish their relative thermodynamic and kinetic stabilities. \cite{polymorphism_pharma_solids_brittain}

\subsection{Crystal Structure Prediction (CSP)}

Despite advances in experimental workflows, no laboratory result can guarantee that every polymorph of a drug has been discovered. 
New forms of even the most thoroughly studied drugs continue to emerge. \cite{Yao2023}
Moreover, experiment requires pure and non-negligible amounts of drug product and can be time- and labor-intensive.
Therefore, the ability to explain or even predict polymorphism from first principles, a task referred to as crystal structure prediction (CSP), is a longstanding scientific goal. \cite{Catlow2023}

The consensus strategy behind most CSP efforts is built on the empirical observation that experimentally observed crystal structures correspond to minima on the free energy (or to an approximation, lattice energy) landscape of crystal structures.
From this observation, CSP can be cast as a multimodal optimization problem, where the goal is to generate all plausible crystal structures of a molecule and rank them by energy.
Experimentally observed polymorphs can be compared against this set of predicted structures, and the experimentalist can easily assess relative stabilities between discovered polymorphs as well as the risk of yet-undiscovered polymorphs.

The difficulty of the CSP task can be ascribed to two orthogonal properties of the problem.
The first is the enormity of the search space.
A CSP protocol must consider both the internal degrees of freedom of the target molecule--drug molecules continue to grow larger and more flexible--as well as the complete space of periodic symmetry operations which define crystal structures. \cite{Agarwal2022}
The second challenging aspect of the CSP problem is the practical cost of ranking the generated polymorphs.
In practice, the energy gaps between potential polymorphs are small, so high-fidelity quantum chemistry methods are needed to accurately discriminate stable polymorphs. \cite{Nyman2015}

\begin{figure}[ht]
  \centering
  \includegraphics[]{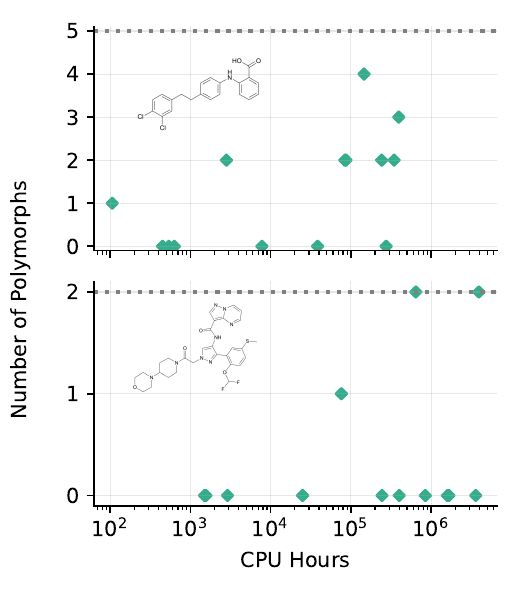}
  \caption{
  Relationship between the success rate (number of successfully predicted experimental polymorphs) and computational cost (reported CPU hour usage) of CSP entries from the CCDC blind challenges for two pharmaceutical targets. Dotted gray lines denote the number of known experimental forms for the molecule at the time of the challenge.
  {\bf Top}: Target XXIII from the sixth blind challenge.
  {\bf Bottom}: Target XXXII from the seventh blind challenge.
  }
  \label{fig:intro_cost}
\end{figure}

The difficulty of CSP is readily apparent in the CSP blind tests organized by the Cambridge Crystallographic Data Centre (CCDC). \cite{lommerse2000first, motherwell2002second, day2005third, day2009fourth, bardwell2011fifth, Reilly2016, Hunnisett2024},
In these tests, participants from academia and industry are provided with the identities of a small number of molecules, and they are tasked with producing ranked sets of plausible crystal structures of each molecule, which are later compared to unpublished experimentally observed polymorphs.
The two most recent blind challenges have each included a pharmaceutical target (XXIII from 6th, XXXII from 7th), and examining performance on these targets is instructive. \cite{Reilly2016, Hunnisett2024} 
Figure \ref{fig:intro_cost} summarizes the performance of the entrants, which is representative of the state-of-the-art, on these two molecules.
From these graphs, a couple of things are apparent. 
First, is that CSP for pharmaceuticals is still a hard, mostly unsolved problem. 
This is evidenced by the fact that a small minority of entrants were able to even generate the experimental polymorphs, let alone rank them well.
The entrants that were successful pay a steep computational cost for their predictions--hundreds of thousands or even millions of CPU hours. 
In many cases, the cost of this compute on commercial clusters may exceed the cost of a full experimental polymorph screen.

\subsection{Replacing Density Functional Theory with Neural Network Potentials}

As discussed earlier, a significant contributor to the cost of CSP is the use of expensive quantum chemistry methods (for ranking and/or force field parameterization), the most common of which is density functional theory (DFT). \cite{Beran2023}
The computational cost of periodic DFT calculations scales as either $\mathcal{O}(N^{3})$ or $\mathcal{O}(N^{4})$ with the size of the unit cell. \cite{martin2020}
This significant cost leads to multiple compromises.
For one, CSP protocols can ultimately generate millions of crystals structures, for which computing DFT energies (let alone optimizing the geometry of each structure with DFT) is impractically expensive.
Instead, it is common to iteratively prune a working set of generated crystal structures with progressively more accurate and expensive computational methods, including simple force fields and semi-empirical potentials like GFN2-xTB. \cite{GFN2}
However, pruning with less accurate methods implicitly introduces potential error into CSP protocols, since less accurate methods are liable to incorrectly understabilize a stable, low-energy crystal structure. 
A second cost-related compromise is the choice of the specific DFT functional.
The most common functional for CSP is dispersion-corrected Perdew–Burke–Ernzerhof (PBE), a robust but imperfect generalized gradient approximation (GGA). \cite{PBE,Reilly2016}
However, there is abundant evidence that this class of semi-local DFT functional can exhibit spurious delocalization, which can compromise lattice energies of molecules such as ROY, axitinib, and galunisertib. \cite{beran2022interplay,Bryenton2022}
More accurate methods, which include non-local hybrid DFT functionals like PBE0, $\omega$B97X-V, and $\omega$B97M-V, are needed to match experiment in many cases. \cite{PBE0,Chai2008wB97,mardirossian2016omegab97m}
Although they provide additional accuracy, these non-local functionals are used sparingly in CSP protocols because of how uniquely expensive exact exchange is for periodic systems. \cite{Hoja2019}

One promising way to alleviate the expensive DFT bottleneck in CSP is with neural network potentials (NNPs), a class of regression models which can be trained on a dataset of structures and reference (DFT) energies and then applied to predict energies of new chemical systems. \cite{Duignan2024, Kocer2022}
In the last decade, NNPs have transitioned from a research curiosity to a practical computational chemistry tool.
NNPs are used in a wide variety of applications, including studying combustion reactions, modeling protein-ligand interactions, catalyst research, and electrolytic chemistry in batteries. \cite{Gomez2024, SabanesZariquiey2024NNPBinding,Yang2022MetadynamicsNNP, Lian2024ODCu, Cao2025ZincElectrolyte, Takamoto2022PFP}

The prospect of achieving DFT-level accuracy at significantly reduced computational cost has made NNPs an attractive option for integration into CSP workflows. As a result, this area has seen considerable activity, including applications to inorganic crystal structure prediction. \cite{Podryabinkin2019, Taylor2024, Omee2024, Rybin2025, Zhang2019}
Some relevant works regarding molecular CSP include those of Day and coworkers, who corrected crystal landscapes from an inexpensive force-field level of theory to PBE. \cite{Butler2024}
Another group developed a CSP protocol around the ANI NNP. 
Although this protocol was both fast and relatively automated, it did not reliably generate experimental crystal structures of many molecules. \cite{Kadan2023}
A team at Schr\"odinger published a CSP protocol and thorough benchmark in which generated crystal structures were pruned with a NNP before DFT re-optimization and re-ranking. \cite{Zhou2025}
Group 20 of the seventh blind challenge used a system-specific version of the AIMNet2 model for each entry. \cite{Hunnisett2024}
For a target CSP molecule, the general AIMNet2 NNP is fine-tuned on `n-mer' energies of a landscape.
The group recently published details in a separate publication. \cite{Anstine2025AIMNet2}

\subsection{Present Contribution}

In this work, we describe a NNP that is specifically architected and trained on DFT computations for the purpose of small-molecule pharmaceutical CSP.
Next, we outline a simple yet robust CSP protocol that makes use of this NNP to both generate and rank crystal structures of pharmaceutical molecules.
This fully automated CSP protocol is tested on a retrospective set of 49 molecules, 47 of which are drugs, which is the largest pharmaceutical CSP benchmark to date.
Lastly, the utility of this fast and automated CSP protocol as a complement to experiment is demonstrated in the practical task of identifying PXRD patterns on a generated landscape.

\section{Lavo-NN}

While there have been promising efforts to apply NNPs to the CSP problem, the full value of this class of predictive models is arguably far from realized.
Many attempts to integrate NNPs into CSP use a general-purpose NNP which may not be ideal for CSP or require fine-tuning a NNP on system-specific quantum chemistry calculations, which is potentially costly. \cite{Zhou2025,Kadan2023,Hunnisett2024}
In contrast, a primary goal in the design of Lavo-NN was the creation of a model that is maximally specialized to the pharmaceutical CSP task (in terms of both training data and model architecture) while generalizing well to new pharmaceutical molecules, so that CSP can be performed on arbitrary pharmaceutical molecules without any re-training on additional system-specific data.

\begin{figure}[ht]
  \includegraphics[]{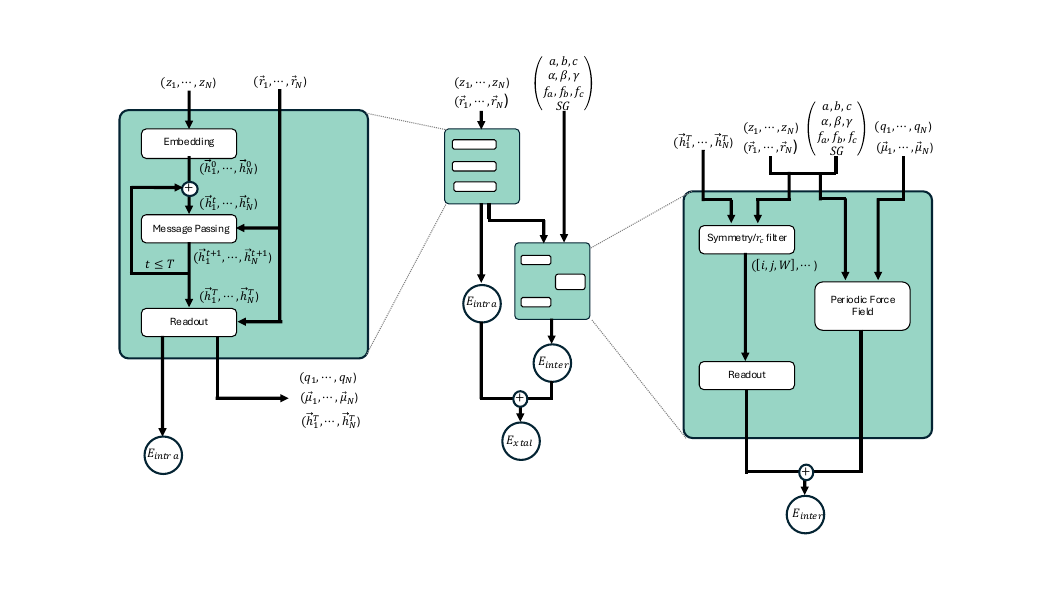}
  \centering
  \caption{Overview of the Lavo-NN architecture. {\bf Center}: high-level representation of the architecture, a NNP designed to efficiently and accurately predict energies and gradients of molecular crystals. {\bf Left}: The intramolecular module predicts the gas-phase energy of the molecule in the crystal as well as atom-in-molecule properties such as atomic charges and dipoles. {\bf Right}: The intermolecular module predicts interaction energies between the molecules in the crystal, correcting a short-range neural term with a long-range periodic force field. Some of the force field parameters are predicted from the intramolecular module.}
  \label{fig:lavo-nn-arch}
\end{figure}

At a high level, Lavo-NN is an equivariant, message-passing NNP.
Like other NNPs of this class, the model computes an energy for a set of periodic atomic coordinates, in which the charge and optional periodic boundary conditions are specified. 
Because Lavo-NN is purposefully designed for use with molecular crystals, an additional assumption can be exploited, which is that a molecular crystal has a defined molecular unit in which atoms are covalently bound.
This precludes the use of Lavo-NN in situations where there is no clear molecular unit (such as materials chemistry or reactive organic chemistry).
The architecture is depicted in Figure \ref{fig:lavo-nn-arch} and a few consequences of this design choice are discussed below.

\subsection{Cost-Effective Intramolecular Message Passing}

Many NNP architectures can be described in the ``message passing" framework. \cite{Gilmer2017}
In this framework, the NNP consists of two phases: a parameterizing message passing phase (in which atoms are iteratively featurized as functions of spatially close neighboring atoms) followed by a readout phase (in which atomic and molecular properties are predicted as a function of these featurizations).
Spatially close neighbors are defined with some interatomic distance cutoff, which is usually on the order of 5-8 $\mathrm{\AA}$.
For solid-state systems like molecular crystals, every atom has many neighbors as defined by this type of cutoff, meaning the message-passing phase can become very expensive, particularly if the message-passing operation involves dense feed-forward layers.
Therefore, for Lavo-NN, the message-passing operation is restricted to intramolecular edges, and intermolecular information can only enter the model through the later readout phase.
This decision was partially motivated by cost, as there are far fewer edges within a molecule than there are in the total crystal.
In addition, intramolecular-only message passing also embodies an interesting inductive bias, which is that an atom's local environment is primarily a function of its intramolecular neighbors, and less of the model's capacity ought to be used in the intermolecular regime.
Lastly, perhaps the most important consequence of constraining message passing to intramolecular edges is that the output crystal energy is decomposable into an intramolecular and intermolecular energy. 
This has repercussions when training the model, and is discussed more in Section \ref{sec:lavo_nn_functional_form}.

\subsection{Physically Motivated Functional Form} \label{sec:lavo_nn_functional_form}

The stability of molecular crystals is governed by a balance of covalent and non-covalent interactions: the most stable polymorphs simultaneously minimize intramolecular strain while forming strong intermolecular interactions.
Relative energies between polymorphs often differ by less than 2 kJ mol$^{-1}$, meaning any potential must be quite accurate across all types of interactions to be actually discriminative in ranking a set of predicted polymorphs. \cite{Nyman2015}
NNPs are well known for accurately modeling covalent and short-range non-covalent interactions; some of the earliest use cases of NNPs were torsion scans and bonded force field parameterization. \cite{ani1, force_balance}
However, traditional NNP architectures can struggle to model long-range non-covalent interactions, as some of us and others have documented. \cite{AP-Net, Ko2021Accounts, Musaelian2023Allegro}
An inability to model long-range interactions is potentially limiting for application to molecular crystals.

%Not data, but functional form.
This deficiency can be attributed to common characteristics of NNP architectures, like small local neighborhoods and atomic partitioning of the density, which do not match the long-range atomic-pairwise nature of non-covalent interactions.
To counteract this, Lavo-NN borrows from the AP-Net architecture, in which readouts are performed over pairs of atoms and augmented with an intermolecular force field, some of the parameters of which are also predicted from a neural network. \cite{AP-Net,AP-Net2}
This architecture is demonstrated to work well for modeling non-covalent interactions in a data-efficient way.

The intermolecular force field of Lavo-NN has the following form:
\begin{equation}
E_{\mathrm{FF}} = E_{\mathrm{FF,2B}} + E_{\mathrm{FF,MB}}.
\end{equation}

$E_{\mathrm{FF,2B}}$ is a simple two-body intermolecular force field consisting of point-charge electrostatics, Born-Mayer exchange, and Tang-Toennies-damped dispersion,

\begin{equation}
    E_{\mathrm{FF,2B}} = \frac{1}{2} \sum_{\substack{i\in M \\ j \notin M}} [ \frac{q_{i}q_{j}}{r_{ij}} + A_{ij} e^{-B_{ij} r_{ij}} - f_{TT}(r_{ij}) \frac{C_{ij}}{r_{ij}^{-6}}].
\end{equation}
\cite{tang1986new}
$E_{\mathrm{FF,MB}}$ is the induced dipole model used in the Amoeba force field \cite{simmonett2015efficient,ponder2002}.
The intermolecular force field is evaluated with a real-space cutoff of 15 $\mathrm{\AA}$.
Periodicity of the crystal is additionally accounted for by applying the Ewald summation method to the electrostatic energy and the electric field of the induced dipole term. \cite{ewald1921berechnung, weenk1977calculation}
Similarly, a tail correction is added to include long-range pairwise dispersion interactions beyond the real-space cutoff.
Force field parameters $A_{ij}$, $B_{ij}$, and $C_{ij}$ are defined by combining rules $A_{ij} = A_{i}A_{j}$, $B_{ij} = \sqrt{B_{i} B_{j}}$, $C_{ij} = \sqrt{C_{i}C_{j}}$, and the corresponding elemental parameters ($A_{i}, B_{i}, C_{i}$) are either borrowed from other force fields or fit to a small number of DFT interaction energy calculations. \cite{schriber2021cliff,grimme2006semiempirical,van2016beyond}

\begin{figure}[ht]
  \includegraphics[width=12cm]{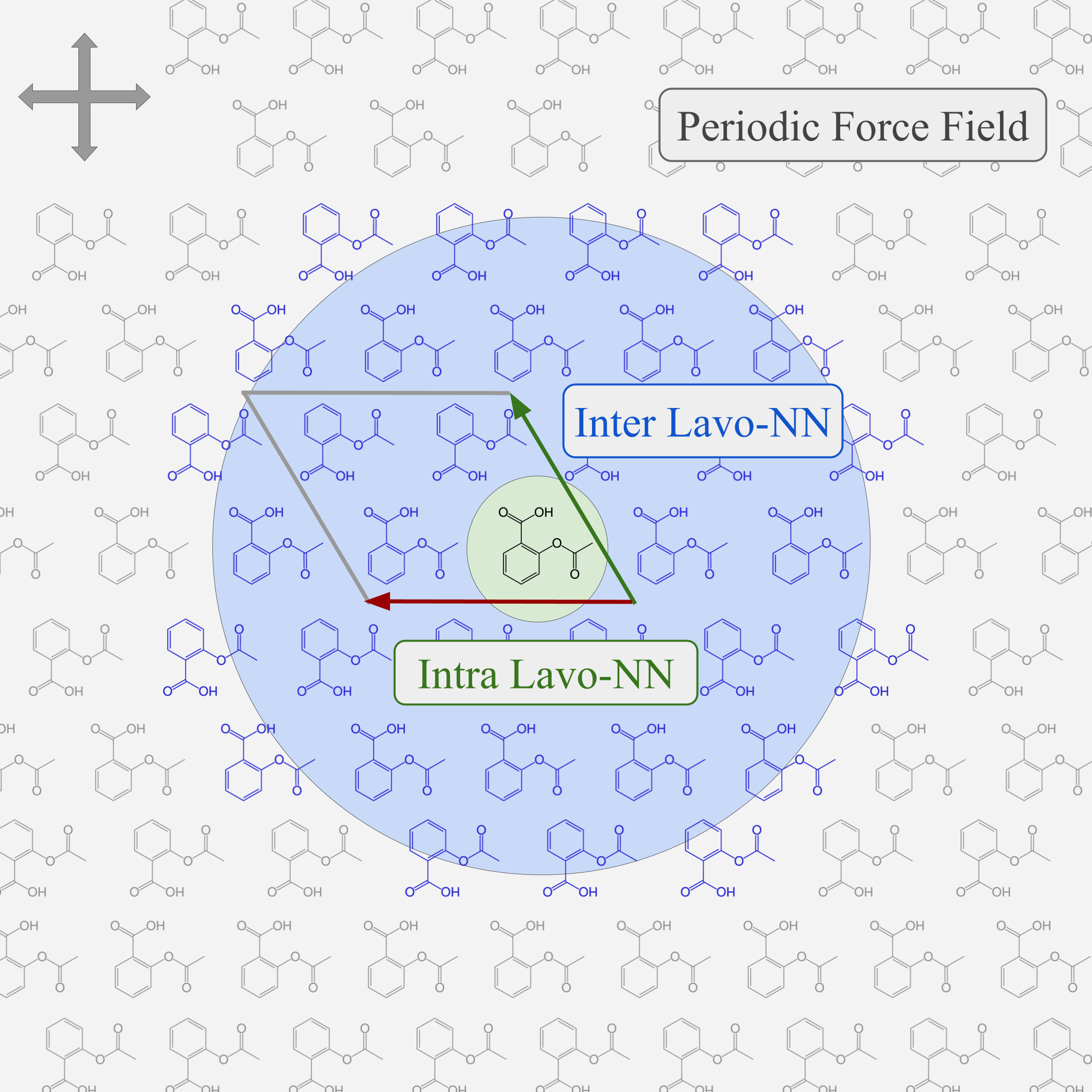}
  \centering
  \caption{Lavo-NN calculates a crystal lattice energy as a combination of an intramolecular energy (green), a sum of intermolecular interactions of close range dimers (blue), and a periodic force field for long range interactions with parameters predicted from the neural network potential (grey).}
  \label{fig:many_body}
\end{figure}

\subsection{Applying the Many Body Expansion}

Lavo-NN is a periodic polarizable force field added to a traditional neural network potential. 
An important consideration is the quantity and nature of the training data used to parameterize the neural component.
Because of the stringent accuracy requirements of CSP and polymorph ranking, it was decided that the model should be trained to a level of theory which is mostly free of delocalization error, meaning hybrid DFT or better. \cite{Perdew2001, Hoja2019,greenwell2020}
However, we are not aware of any large datasets of hybrid periodic DFT calculations, and developing a sufficiently large dataset at this level of theory is prohibitively expensive.

Instead of explicitly training the Lavo-NN model to predict the total energy of a crystal, an attractive alternative strategy is to train a model to predict crystal energies via the truncated many-body expansion (MBE).
In the MBE, the energy of a periodic crystal can be decomposed in a series of aperiodic N-body contributions:
\begin{equation}
E_{xtal} \equiv E_{0} + \frac{1}{2}\sum_{I > 0} \Delta E_{0I} + \frac{1}{3}\sum_{I > J > 0} \Delta E_{0IJ} + \cdots,
\end{equation}
where the one-body term, $E_{0}$ is the isolated energy of the unique monomer, the two-body term consists of dimer interaction energies ($\Delta E_{0I}$) between the unique monomer and some other monomer $I$ in the crystal, the three-body term consists of analogous non-additive trimer interaction energies ($\Delta E_{0IJ}$), and so on.
Application of the MBE to molecular crystals is well-studied by the groups of Sherrill, Beran, and others. \cite{Borca2023MBE, Kennedy2014ThreeBody, Wen2011FragmentQM, Herman2023IceMBE, beran2010predicting,sargent2023benchmarking}
The MBE is also the basis for some CSP protocols. \cite{mcdonagh2019machine,Bannan2025CSPCloud}

The neural component of Lavo-NN is trained to the MBE, truncated at second order:
\begin{equation}
   E_{xtal} \approx E_{0} + \frac{1}{2}\sum_{I > 0} \Delta E_{0I},
\end{equation}
where the intramolecular and intermolecular modules of the model are separately trained to predict the first and second order terms of this expansion.
This MBE-based training task therefore requires two datasets: a traditional total energy dataset with which to train the intramolecular module, and a dimer interaction energy dataset with which to train the intermolecular module.
The first dataset should also contain atomic properties, so that the intramolecular module can be used to predict conformation-dependent atomic properties for the intermolecular module's force field.
This MBE-based training yields a number of benefits.
One obvious benefit is that the associated training data (monomer energies and charges; dimer interaction energies) is more plentiful and generating additional data is more affordable when targeting high-accuracy hybrid DFT.
Not only is data easier to procure, it is also easier to customize.
The size, composition, and even the level of theory of both datasets are decoupled and can be customized for the particular task.
One might use a higher-accuracy DFT functional or larger basis for the intramolecular data, and be able to get by with a less expensive level of theory on the intermolecular data.
Lastly, by explicitly decomposing the task of predicting a crystal energy into physically meaningful sub-tasks of predicting molecule and dimer interaction energies, it is plausible that this physical inductive bias might enable the model to learn more efficiently.

It is also worth commenting briefly on errors associated with second-order MBE truncation.
Research has shown that higher-order terms in the MBE contribute a small but sometimes non-trivial amount of energy to $E_{xtal}$. \cite{nelson2024convergence,liang2023can}
In our experience, while higher order MBE terms can be non-zero, they usually cancel between polymorphs of the same molecule, so they can be disregarded for the CSP task, where relative stabilities of polymorphs is of interest.
Additionally, Lavo-NN's polarizable force field captures polarization interactions to infinite order in the MBE. 
The missing energy associated with this approach is therefore mostly attributed to many-body dispersion, which is liable to cancel between polymorphs. \cite{otero2020many}

\subsection{Training Data}

The Lavo-NN model described above was trained on a uniquely constructed dataset which was tailor-made for the task of performing CSP on drugs and drug-like molecules.
This dataset was developed in an iterative, self-consistent fashion: the current best Lavo-NN model was used to generate low-energy crystal structures, which served as additional training data for the next iteration of the model.

The starting point for this training procedure was a Lavo-NN model trained on two pre-existing public datasets:
The intramolecular module was trained on SPICE v2.0, and the interaction module was trained on the Splinter dataset. \cite{Eastman2023, Spronk2023}
Next, an assortment of 10k organic molecules was randomly selected from the curated set of SMILES strings associated with the SPICE dataset.
For each of the 10k SMILES strings, a simplified version of the full CSP protocol described in Section \ref{sec:csp_protocol} was performed, using the initial Lavo-NN model to generate and optimize $\sim$100 crystal structures per molecule.

From the resulting Lavo-NN-optimized 1M crystal structures, molecular conformers and closely interacting dimers were extracted, and in some cases rattled and/or fragmented.
Fragmentation was performed to limit the size of the individual molecules and dimers, and therefore the cost of computing DFT labels.
In all cases, only single bonds were fragmented and capped with hydrogen atoms, and the hydrogen positions were optimized.
DFT energies, gradients, and densities were computed with the \PSIfour\ program. \cite{Smith2020Psi4}
A custom minimal-basis iterative stockholder (MBIS) routine was developed to extract atomic multipoles from the \PSIfour\ densities. \cite{verstraelen2016minimal}
For the intramolecular dataset, the $\omega$B97X-D3(BJ)/def2-TZPPD level of theory was used, consistent with the SPICE dataset. \cite{Chai2008wB97, Grimme2010D3, Grimme2011BJ, Weigend2005def2, Hellweg2015TZVPPD}
For the intermolecular dataset, the less expensive composite $\omega$B97X-3c method was chosen, and a fraction of the Splinter dataset was recomputed at this level of theory. \cite{Mueller2023wB97X3c}
In internal benchmarks, it was found that this method provides a good cost-accuracy tradeoff, and its allowance of monomer-centered bases for the interaction energy calculation makes running this method economical.
The initial Lavo-NN model was trained on this new pair of datasets, and the whole procedure was repeated for an additional five iterations, at which point the accuracy of the model plateaued.

\subsection{Implementation Details}

Lavo-NN is efficiently implemented in a combination of C++, CUDA, and Python.
The force field is written in both C++ and CUDA and CPU and GPU execution, respectively.
In both implementations, computation of the force field energy ($E_{FF}$) and gradient is fused to minimize redundant work.
The neural component of the crystal energy is written in PyTorch, which supports multiple hardware backends.
Gradients of neural components are computed with PyTorch's `autograd' feature.
Gradients are taken with respect to the atomic positions of the unique atoms in the unit cell, as well as the three unit cell lengths and three unit cell angles, unless constrained by the symmetry.

Relative to other NNPs, significant efficiency gains are realized in the exploitation of space group symmetries.
The periodic symmetry of any molecular crystal can be described by one of 230 space groups, which among other things, determines the number of symmetry-redundant molecules per unit-cell. \cite{IUCR2016}
In some of the most common space groups (P$\bar{1}$, P$2_{1}$/c, C2/c) there are 2, 4, or even 8 copies of the same asymmetric unit (molecule) per unit cell. \cite{Groom2016}
A naive NNP implementation would require computing the features of each symmetry-equivalent molecule.
In Lavo-NN, symmetry information is integrated such that both the message passing and readout is performed over only unique atoms (or unique pairs of atoms), further reducing cost.
This is purely an engineering optimization; specific features and readouts are identical to the naive implementation, so accuracy is unaffected.

\section{Crystal Structure Prediction Protocol}  \label{sec:csp_protocol}

The multimodal CSP task requires generating all low-energy crystal structures of a target molecule, and doing so as efficiently as possible. For small, rigid molecules brute force solutions are feasible, but the large flexible molecules common in drug development require carefully designed sampling strategies to efficiently explore all intramolecular and crystallographic degrees of freedom. 
Here, we describe a CSP protocol developed for this task.
It is designed to be efficient, parallelizable, and largely automated, requiring minimal manual input.
Specifically, the only inputs are a SMILES string of the target molecule and, if the molecule is chiral, whether the crystal should be racemic or enantiopure.
This version of the protocol targets molecular crystals with one unique molecule in the periodic unit cell ($Z' = 1$); future versions will address more complex crystal structures. 
The protocol is built around the Lavo-NN potential.
This removes the need for potentially expensive force field parameterization in the structure generation step, as is common in many CSP protocols. \cite{Neumann2008}
The accuracy of this potential also allows the user to optionally skip final re-ranking with DFT.
If DFT is run after structure generation, the Lavo-NN generated structures are accurate enough that only single-point energy calculations with DFT are required.
This is preferable to DFT optimizations, which require tens of sequential energy and gradient calculations per structure.

\subsection{Conformer Generation}
Generating reasonable conformers of the target molecule is a common subtask within CSP, and is a component of this CSP protocol.
Some protocols require user-defined conformer sets or perform expensive DFT calculations to explore intramolecular degrees of freedom.
Here, a simpler approach is used.
RDKit is used to process the initial SMILES string, and the connectivity graph is used to determine the most flexible degrees of freedom (free torsions, flexible rings, and invertible nitrogens). \cite{rdkit}
Random conformers are then generated by uniformly sampling these degrees of freedom, under the constraint that the molecular graph is not broken.
Other degrees of freedom (stretches and bends) are not explicitly sampled but rather are implicitly sampled by later optimizing the crystal structures containing these conformers with the Lavo-NN potential.

\subsection{Geometric Crystal Generation} \label{ssec:geometric_generation}
A $Z' = 1$ crystal can be defined as a conformer placed at a specific position and orientation in a periodic unit cell.
This adds up to 13 additional degrees of freedom to the simpler conformer generation problem: three translational, three rotational, six defining the unit cell geometry, and an additional discrete variable, the space group, which defines the symmetry of the unit cell.
Depending on the space group symmetry, some of the unit cell parameters can be constrained.

For large, flexible molecules, the interdependence of these variables makes obtaining realistic, densely packed crystals difficult; naively sampling these variables results in crystals that either contain unphysical steric clashes or extremely low densities.
Here, crystal structures are instead generated with a bottom-up, geometric approach.
Given a conformer and a space group, the composite symmetry operations of the space group (translations, points of inversion, screw axes, glide planes, etc.) are applied in random order and orientation to the conformer, with the symmetry element placed as close as possible to the conformer without causing a steric clash.
Finally, the entire crystal is optimized with Lavo-NN.

\subsection{Monte Carlo Crystal Generation} \label{ssec:monte_carlo}
Using geometric generation described in Section \ref{ssec:geometric_generation}, a random, densely packed crystal can be easily generated given a conformer.
However, while these generated crystals are reasonable crystal structures and local minimum, the size of the search space makes it difficult for this generation strategy to quickly find the lowest energy crystals, particularly for flexible molecules with many internal degrees of freedom.

This problem is addressed by performing a semi-local Monte Carlo search across the crystal structure landscape for each geometrically generated crystal structure.
A Monte Carlo search requires a defined move set, which is chosen to be singular changes to the flexible internal coordinates of the crystal's conformer (free torsion, invertible nitrogen, or flexible ring) followed by a geometry optimization with the Lavo-NN potential.
Because of the dense packing of atoms in these crystal structures, directly changing internal degrees of freedom like torsions almost always results in unphysical steric clashes in the crystal structure. 
To make these Monte Carlo moves more realistic, the internal coordinate change of each move is applied using an interpolative procedure.
An interpolation in Cartesian coordinates is performed between the initial and final conformer, and at each intramolecular interpolant, the crystal is re-optimized using an inexpensive force field with the conformer's internal geometry held rigid.
This allows the conformer to translate and rotate, effectively moving out of the way to accommodate the new conformation while preserving a dense crystal packing motif.

A sequence of Monte Carlo moves is applied per crystal generated with the geometric generation protocol, and each move is accepted according to the Metropolis condition. \cite{li1987monte}
Each crystal structure (accepted or rejected) is added to a total working set of all generated crystal structures.

\subsection{Convergence Assessment} \label{ssec:convergence_assessment}

One difficulty of crystal structure generation is that no clear-cut method exists to determine exactly how many crystal structures should be generated.
Terminating the generation step too early runs the risk of missing important structures, while performing excessive structure generation incurs unnecessary computational cost.
In this CSP protocol, assessing convergence is slightly simplified by separately and independently determining convergence per space group.
This is motivated by the empirical observation that some space groups simply produce fewer crystal structures than others, and therefore require less sampling.
An example of a fast-converging space group is $P1$, which only has three degenerate translation operators.
Not only do some space groups converge faster than others, but some space groups are less likely to produce low-energy crystal structures on account of their symmetry operations.
Space group $Pmmm$ is one such space group, in which real-life crystals are rarely found and computationally generated crystal structures are almost exclusively high in energy. \cite{CCDC_CSD_2025}
This space group is likely disfavored due to the three mirror planes, which both prevent dense packing and encourage unfavorable electrostatic interactions. \cite{brock1994towards}
Assessing convergence on a per-space-group basis therefore allows more compute to be allocated to space groups in which there is a better chance of generating low-energy structures.

Convergence assessment can be cast as estimating the probability that a low-energy crystal structure exists in this space group and has not yet been generated.
As long as this probability remains non-negligible, additional crystal structures are generated in the space group.
Of course, there is no \textit{a priori} way to assess which or how many low-energy crystal structures have not been generated yet; finding these structures is the problem at hand.
However, the number of unseen crystal structures can be estimated by examining the distribution and counts of found crystal structures.
The more times each already-discovered crystal structure is independently generated, the less likely it is for an unseen crystal structure to exist.
This framing is surprisingly similar to the well-studied species abundance modeling problem in ecology, where one attempts to estimate the biodiversity of an area by sampling individual organisms and tallying unique species. \cite{colwell1994estimating}
In this analogy, each ``species" is a unique crystal structure, and each ``organism" is a generated crystal structure.
We found that the Poisson-lognormal distribution works well for modeling counts of unique crystals:
\begin{equation}
P(N=y)
=
\int_{0}^{\infty}
\frac{e^{-\lambda}\,\lambda^{y}}{y!}
\;\times\;
\frac{1}{\lambda\,\sigma\sqrt{2\pi}}
\exp\!\Bigl[-\frac{(\ln\lambda-\mu)^2}{2\sigma^2}\Bigr]
\;d\lambda.
\end{equation}
The parameters of the distribution $\mu$ and $\sigma$ are continually fit to best reproduce the empirical frequencies of unique crystal counts ($f_{1}, f_{2}, \cdots, f_{y})$ of each space group by minimizing the Kullback–Leibler divergence between the empirical and model distribution, subject to the fact that $f_{0}$ is unknown.
Finally, the product of $P(N = 0)$ and the number of unique crystals in the space group, which is the estimated number of undiscovered crystals, is used as the convergence metric.

In practice, as crystal structures are generated, only the lowest energy structures are maintained.
For each newly generated structure, a similarity comparison to each member in the working set of existing structures is made with a custom implementation of the RMSD$_{N}$ metric.
This works by extracting and aligning spherical clusters of $N$ molecules from a pair of crystals, and their geometric similarity is quantified via root mean squared deviation (RMSD). 
This protocol defaults to a cluster size of 20 molecules and a heavy-atom-only similarity cutoff of 0.8 $\mathrm{\AA}$. \cite{chisholm2005compack}
If a newly generated structure is a match to any existing structure by the RMSD$_{20}$ metric, only one of the two is kept (that with lowest energy), and the times-generated count associated with the unique crystal is incremented.
%It is important to increment the tally only when the discovered crystal is generated from an independent Monte Carlo simulation.

\subsection{DFT Re-Ranking}\label{ssec:dft_reranking}

Once convergence of the structure generation step has been reached in all space groups, the protocol could be considered complete.
In many cases, the resulting landscape of crystal structures is sufficient to answer questions about polymorphism.
In some situations, however, high-accuracy DFT rankings are desirable.
Therefore, by default, the lowest energy structures of the generated landscape are re-ranked with periodic PBE-D3(BJ), plus a monomer PBE0 correction.
The periodic PBE-D3(BJ) calculations are performed in Quantum Espresso and the monomer PBE0 corrections are performed with Psi4. \cite{giannozzi2020quantum, Smith2020Psi4}
This pragmatic level of theory avoids the cost of hybrid periodic DFT while still eliminating much of the intramolecular delocalization error via the hybrid monomer correction. \cite{greenwell2020, Bryenton2022}

\section{Software Architecture for Crystal Structure Prediction} \label{sec:software_architecture}
Section \ref{sec:csp_protocol} defines a CSP protocol for generating a set of predicted polymorphs for an arbitrary pharmaceutical molecule.
Here, a software architecture is described that implements this protocol according to three core design goals: simplicity, high-throughput, and cost-effective scalability. 
Deployed on the Amazon Web Services (AWS) cloud platform, the architecture uses a decoupled model in which a central coordinator orchestrates large-scale simulations across a fleet of distributed, containerized workers. \cite{aws}
\begin{figure}[ht]
  % pull left by 0.1\textwidth and oversize a bit
  \noindent\hspace*{-0.0\textwidth}%
  \includegraphics[width=1.0\textwidth]{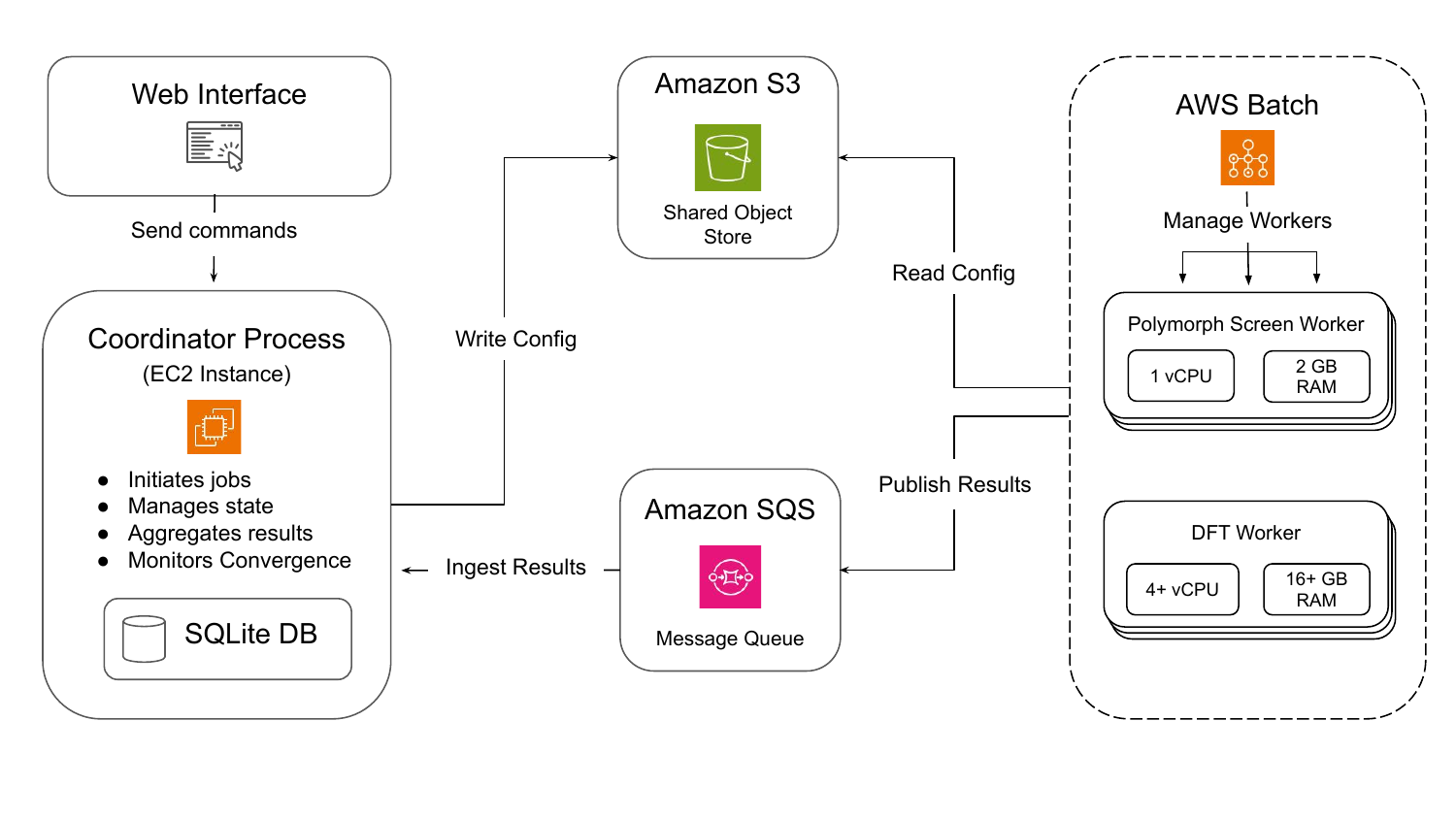}
  \caption{Visual overview of the software architecture. The coordinator process initiates jobs and manages the job state. The coordinator communicates with the distributed workers through S3 storage buckets and an SQS message queue. AWS Batch deploys thousands of distributed workers that independently generate millions of possible polymorphs. The system is managed and controlled from a web interface and is deployed on AWS.}
  \label{fig:software_arch}
\end{figure}
\subsection{Coordinator Process}
The CSP protocol is controlled by a single process on a central EC2 virtual machine, referred to as the ``coordinator''.  
The coordinator initiates the Monte Carlo structure generation step described in Section \ref{ssec:monte_carlo} by writing global configuration parameters, such as space group selection and molecule configuration, to a shared object store on Amazon S3. \cite{aws-s3} The coordinator process then initializes a set of distributed worker processes via AWS Batch and sets up a dedicated Amazon SQS message queue to track crystal structures generated by these worker processes. \cite{aws-sqs}

Once the worker processes are launched by AWS Batch, the coordinator monitors the message queue for new crystal structures generated by the distributed worker processes. \cite{aws-sqs} 
The coordinator maintains a refined set of unique low-energy crystal structures by regularly performing the convergence assessment described in Section \ref{ssec:convergence_assessment}.
When a particular space group is deemed converged, the coordinator communicates this information to the workers by updating configuration data in the S3 object store associated with the particular CSP task.
After the generation task is finished, the coordinator initiates the optional DFT re-ranking step in a process analogous to the Monte Carlo structure generation.
DFT re-ranking uses a different class of worker process, as the CPU, memory, and storage demands of DFT calculations are quite different from those of the structure generation task.

State management of the current status of the protocol is handled by a SQLite database which is local to the coordinator process. \cite{sqlite2025hipp} 
This choice avoids the complexity of a networked database system while achieving high performance.
To ensure fault tolerance for this central component, the system performs periodic and automated snapshots of the server's underlying storage volume. 

\subsection{Distributed Worker Processes}

To achieve cost-effectiveness, the methodology favors horizontal scalability across large numbers of small, commodity compute instances, which offer the most favorable cost-per-core. 
The architecture's scalability and simplicity are achieved by deploying the CSP workers using containers on a managed cloud service. 
Docker is used to package the application, including the operating system and pinned versions of all scientific libraries, into a single, portable container image. \cite{docker} 
This guarantees that calculations are fully reproducible and agnostic to the specific compute resources of the underlying hardware.
These container images are then executed using AWS Batch, a fully managed service that automates the deployment, scheduling, and management of thousands of concurrent tasks. \cite{aws-batch} 
This combination offloads the complexity of provisioning and scaling a large compute cluster to AWS, providing a simple and robust mechanism to achieve massive horizontal scaling.

The core task of the protocol, Monte Carlo crystal generation, is well suited to this model as it is embarrassingly parallel, allowing thousands of ephemeral workers to explore the conformational space independently without inter-process communication. \cite{heath1986hypercube, foster1995designing} 
Each worker is a lightweight process, typically running on a single CPU core. This small footprint enables fast start times of less than a minute and throughput of thousands of concurrent cores across a wide range of hardware. 
The lifecycle of a worker is straightforward: upon startup, it retrieves global simulation parameters from S3, performs crystal generation in its assigned space group, and publishes any identified low-energy candidate structures to the SQS queue for subsequent ingestion by the coordinator.

The flexibility of the architecture, derived from its modular and container-based design, allows it to be easily adapted to run the resource-intensive DFT calculations in later stages of the protocol. 
For these tasks, workers are configured to request instances with more significant resources (e.g., 4-16 vCPUs and 16-128 GB of memory) and, where necessary, high-throughput local NVMe SSD storage, all while using the same underlying orchestration and workflow model.

\subsection{Cost Optimization through Fault Tolerance }

Cloud platforms like AWS enable dynamic scaling of compute resources, creating variable spare capacity in datacenters.
To manage this variability, cloud providers offer ``spot'' instances — discounted virtual machines that come with the tradeoff of fluctuating pricing and availability that can be terminated with minimal notice. \cite{LIN2022103718}
As a cost-saving measure, the CSP protocol makes extensive use of these spot instances. 
Consequently, the worker processes that run on spot instances were implemented to be robust against sudden instance termination. 
The Monte Carlo structure generation phase operates incrementally with frequent checkpointing, ensuring each computational unit completes within minutes. 
When a spot instance is terminated, minimal work is lost due to this fine-grained checkpointing strategy.
AWS Batch automatically replaces terminated instances with new workers that resume calculations from the most recent checkpoint, preserving computational progress and making spot instances practical for the workload.
The price and availability of the spot instance correlate with the size of the virtual machine. Larger instances typically cost more and have limited supply. This aligned well with the lightweight and highly parallel design of the Monte Carlo sampling algorithm, so it was most cost effective to provision large numbers of single-core instances to get access to the greatest availability of compute and lowest prices.

\section{Validation / Results}
\begin{figure}[!ht]
  \includegraphics[width=0.94\textwidth]{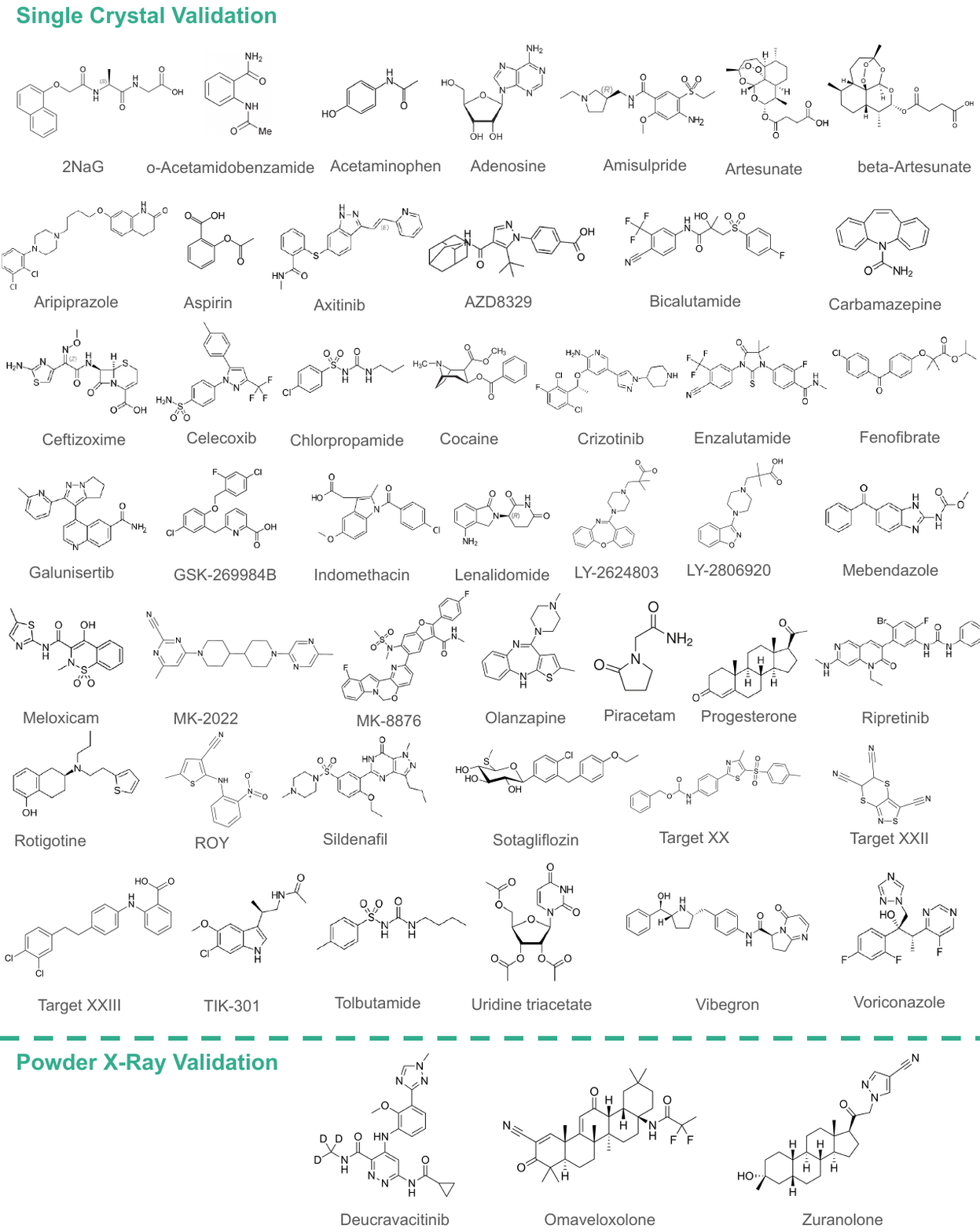}
  \caption{Benchmark molecules used to assess the CSP protocol. \textbf{Top}: 46 molecules (of which 44 are pharmaceuticals) with one or more experimental crystal structure in the CSD. \textbf{Bottom}: Three current drug molecules for which no experimental crystal structures have been published, but PXRD patterns are available.}
  \label{fig:molecules}
\end{figure}

CSP protocols are evaluated on how well they match experimental measurements. 
Following the standard of the CCDC blind challenges, we collected a benchmark of molecules and corresponding experimental crystal data with which this protocol and others can be assessed.
This set of 49 molecules, summarized in Figure \ref{fig:molecules}, is associated with 110 experimentally observed polymorphic forms.
Although there are other large CSP benchmarks, 47 of the 49 molecules here are drugs or associated with some form of biological activity, making this the largest benchmark of pharmaceutical molecules. \cite{Zhou2025,taylor2025predictive} 
The two non-pharmaceutical molecules are ROY, known because of its record number of polymorphs, and Target XXII from the sixth blind challenge. \cite{beran2022many,Reilly2016}
They were included because they have been thoroughly studied, both experimentally and computationally, and thereby act as a useful reference.
Three of the molecules in the benchmark have polymorphs which are only characterized by powder patterns, which is representative of common working conditions (see Section \ref{ssec:pxrd_benchmark}).
%The molecules used to validate the CSP protocol were chosen for the amount and quality of experimental data for their solid forms, their flexibility, and diversity, with a preference for drug-likeness. 
%9 unique molecules associated with 110 stable, experimentally-validated polymorphic forms were chosen. 
%Of the 49 molecules, three have only PXRD data available, to which we refine simulated patterns in search of the experimental form. 
%For the remaining 46, we compare to curated single-crystal X-ray diffraction (SCXRD) structural data from the Cambridge Structural Database (CSD). 
%Diagrams of all molecules used to validate the CSP protocol are shown in Figure \ref{fig:molecules}.
%These include molecules such as ROY, aspirin, acetaminophen, o-acetamidobenzamide, Target XXII from the CCDC’s sixth blind test, and progesterone. 
%The validation dataset furthermore emphasizes polymorphic systems when possible in order to probe the ability of neural network and DFT energy methods to determine the relative stability of polymorphs.
%Of the 49 molecules studied in total, 22 are associated with multiple polymorphs.

Polymorphs in the benchmark were limited to $Z' = 1$ crystal structures, in line with the current CSP protocol.
In cases where multiple CSD entries exist and correspond to the same polymorph, the most reliable one is chosen, preferring the one with the smallest R-factor. \cite{Urzhumtseva2009}
As an added test of the robustness of the protocol, molecules which crystallize in less common space groups were intentionally included.
These molecules include chlorpropamide and tolbutamide, which have polymorphs in space group Pbcn (\#60), sildenafil with a polymorph in Pccn (\#56), and finally the well-known example of rotigotine, with a polymorph in P4\textsubscript{3} (\#78). \cite{chlorpropamide, tolbutamide, sildenafil, rotigotine_form_1} 
\newpage

In all instances of the CSP protocol, crystal structure generation was restricted to a subset of 49 of the most common space groups that in total cover 97\% of the crystals in the CSD. \cite{CCDC_CSD_2025}
For chiral molecules, the selection of space groups was constrained to the Sohncke (or in some cases, non-Sohncke) space groups when appropriate.
These choices correspond to common practices in CSP, where the focus is on the most probable space groups and enantiopurity is often controlled during synthesis. 

To best represent a real-world solid form assessment, two molecules, mebendazole and progesterone required additional care in configuring the CSP protocol.
Mebendazole tautomerizes about the benzamidizole and neighboring amine, yielding three distinct tautomers for which CSP was separately performed, and each of which is associated with a unique crystal structure. \cite{Bravetti2022Mebendazole, perry25}
Endogenous, human progesterone is an enantiomer termed “natural” or “nat” progesterone, and has two known polymorphic forms that crystallize in the typical situation of enantiopure starting material. \cite{original_progesterone}
However, a more stable racemate crystallizes when the other enantiomer “ent-progesterone” is present. \cite{racemic_progesterone} 
Therefore, the CSP protocol was executed in only the Sohncke space groups to reflect enantiopure starting material, and again in the non-Sohncke space groups.

\subsection{Retrospective CSD Benchmark} \label{ssec:benchmark}

% Some text on the overpopulation problem. Can it apply here?

%Another challenge limiting the impact of CSP is the overpopulation problem, the fact that theoretical CSP predicts a very large number of plausible low-energy crystals. 
%This is due to the rugged potential energy landscape, which is largely unaffected by the use of improved potentials. 
%Overpopulation comes from a combination of factors: some energetic minima are thermodynamically equivalent and will freely interconvert, some reliably and irreversibly convert to a more stable form, and some are kinetically inaccessible under normal conditions. 
%This hinders the value of CSP, as the existence of metastable polymorphs cannot generally be predicted, only verified.

\begin{figure}[!ht]
  \includegraphics[width=\textwidth]{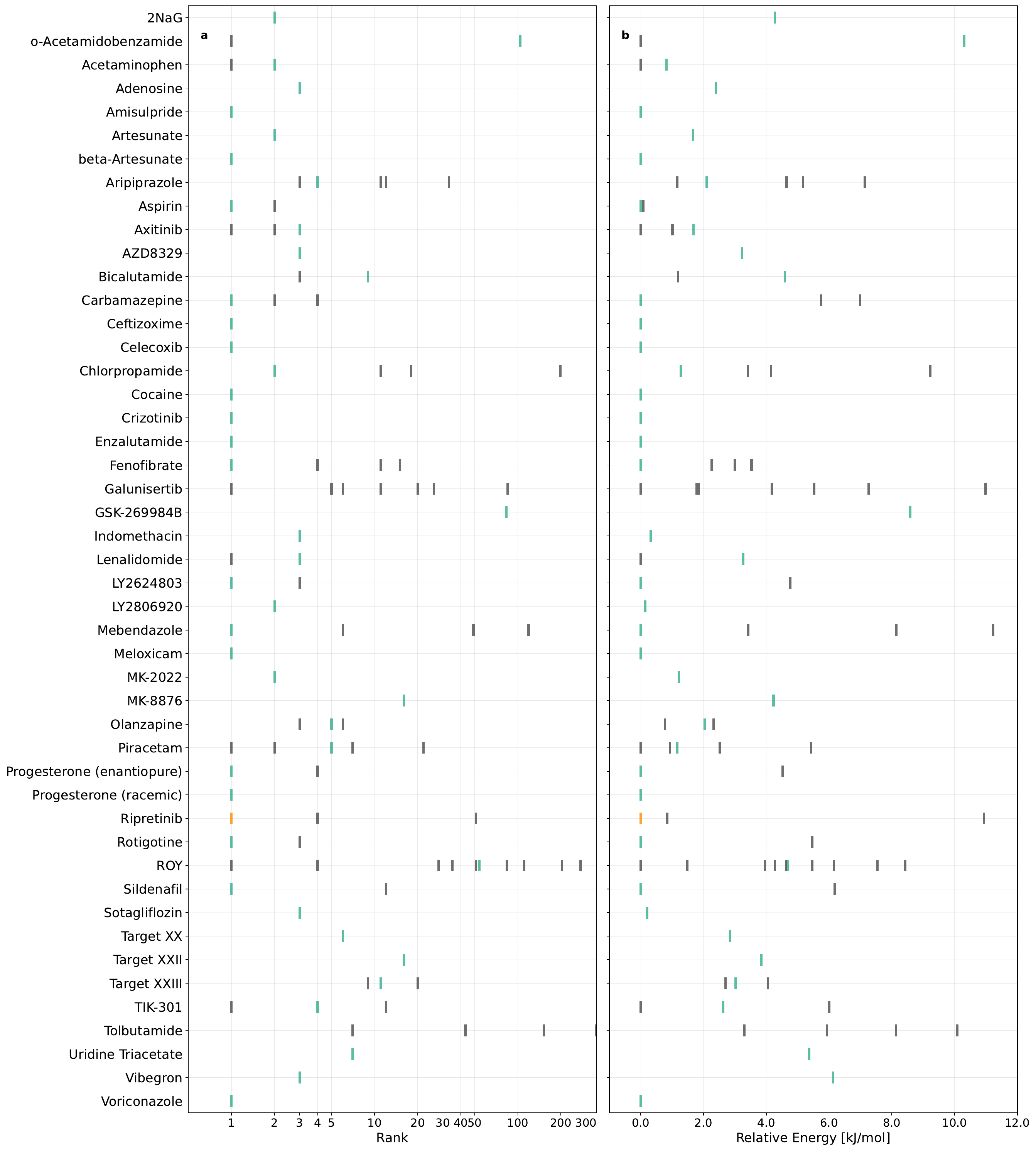}
  \caption{Rankings (A) and relative energies (B) of all experimental crystals found in the study. The experimental thermodynamically most stable polymorph is marked in green. For ripretinib, the most stable experimental polymorph is $Z' = 2$, so this structure was manually added to the structure list and optimized and ranked using the standard CSP workflow. This structure is marked in orange.}
  \label{fig:ranking_results}
\end{figure}

The objective of the CSD benchmark portion of this study was to apply the CSP protocol described in Section \ref{sec:csp_protocol} to 46 molecules, for which 104 distinct, experimentally determined crystal forms exist, sourced from the Cambridge Structural Database (CSD). \cite{Groom2016} The protocol is assessed on its ability to generate the experimental crystal structures as well as to rank each as low-energy relative to other generated forms. When a molecule is polymorphic, an additional goal is accurate prediction of the relative stabilities between forms. As shown in Figure \ref{fig:ranking_results}, the protocol is successful. All 104 polymorphs are both generated and ranked near the bottom of the corresponding crystal structure landscape. 
One polymorph, galunisertib Form I, is excluded from analysis, as it is a known highly metastable and hygroscopic form that has not been sampled in other CSP efforts. 
This polymorph is thought to be as much as 30 kJ mol$^{-1}$ less stable than the most stable form. \cite{Bhardwaj2019Galunisertib}

The results demonstrate the efficiency of the protocol in identifying thermodynamically stable polymorphs.
As illustrated in Figure \ref{fig:ranking_results}, which plots the relative rankings and energies of the identified experimental structures, our protocol successfully predicted the known thermodynamically most stable crystal as the global minimum for 18 of the 46 molecules.
In an additional 8 final landscapes, the experimentally observed stable form was predicted to be within 2 kJ mol$^{-1}$ of the global minimum, a deviation that falls within the expected error margin for the DFT functional used in the final re-ranking stage.
Furthermore, the protocol consistently ranked known experimental polymorphs among the most energetically favorable candidates. Across the entire benchmark, 87\% of the candidate structures that matched known experimental forms were ranked within the top 50 predicted structures. This indicates a high success rate for placing experimentally relevant structures within a small, low-energy window for further consideration.

It is important to assess not just the accuracy but also the cost of any CSP protocol. 
Given enough compute power, even the most inefficient CSP protocol will be successful.
The total computational cost for this benchmark amounted to approximately 438k CPU hours, a dramatic reduction in the typical the computational cost of such an effort. 
These CSP tasks were conducted in just days of wall-time by a single person using highly automated infrastructure, which is described in Section \ref{sec:software_architecture}. 
Table {\ref{tab:cpuhours}} provides a detailed breakdown of the CPU cost across the three main stages of the protocol: the initial Monte Carlo structure generation, the coarse re-ranking with PBE-D3, and the final ranking incorporating a PBE0 monomer correction. 
The table further illustrates the breakdown of CPU usage across each stage, with DFT-based re-ranking consuming a comparatively small fraction of the total resources. 
This cost structure demonstrates reasonable scalability with molecule size and confirms that the approach significantly reduces the reliance on expensive DFT calculations, a traditional bottleneck in CSP workflows.
The average CSP protocol in this study costs 8.4k CPU hours, with a median cost of only 5.4k CPU hours.
This low computational cost opens the door to routine use of CSP in the early stages of drug discovery such as lead optimization, where development concerns can be added to the decision-making process in a way that was not feasible previously. \cite{Hughes2011}

\newpage

\begin{table}[htbp]
\centering
\scriptsize
\caption{
Per-molecule cost (in CPU hours) of each CSP performed in the benchmark. The steps are split into structure generation, single-point periodic PBE-D3(BJ) energy assessment on the lowest-energy structures, and a monomer PBE0 correction on the lowest-energy structures.}
\vspace{0.5cm}
\label{tab:cpuhours}
\begin{tabular}{l r r r r}
\toprule
\textbf{Compound} & \textbf{\shortstack{Structure\\Generation}} & \textbf{\shortstack{Periodic\\PBE-D3(BJ)}} & \textbf{\shortstack{Monomer\\$\Delta$PBE0}} & \textbf{\shortstack{Total}} \\
\midrule
2NaG & 5.1k & 302 & 137 & 5.6k \\
o-Acetamidobenzamide & 562 & 888 & 52 & 1.5k \\
Acetaminophen & 717 & 544 & 80 & 1.3k \\
Adenosine & 426 & 286 & 31 & 743 \\
Amisulpride & 13.9k & 3.0k & 214 & 17.1k \\
Artesunate & 4.0k & 439 & 133 & 4.6k \\
beta-Artesunate & 5.6k & 215 & 123 & 5.9k \\
Aripiprazole & 33.8k & 870 & 404 & 35.1k \\
Aspirin & 762 & 1.1k & 77 & 2.0k \\
Axitinib & 3.1k & 95 & 40 & 3.2k \\
AZD8329 & 2.5k & 122 & 81 & 2.7k \\
Bicalutamide & 3.7k & 460 & 98 & 4.3k \\
Carbamazepine & 1.0k & 1.3k & 37 & 2.4k \\
Ceftizoxime & 1.2k & 101 & 48 & 1.4k \\
Celecoxib & 3.8k & 2.2k & 124 & 6.1k \\
Chlorpropamide & 2.5k & 2.5k & 228 & 5.2k \\
Cocaine & 2.0k & 1.7k & 126 & 3.8k \\
Crizotinib & 8.7k & 858 & 119 & 9.7k \\
Enzalutamide & 4.2k & 4.8k & 266 & 9.2k \\
Fenofibrate & 7.2k & 12.4k & 1.1k & 20.7k \\
Galunisertib & 2.5k & 1.2k & 158 & 3.8k \\
GSk-269984B & 15.7k & 6.3k & 527 & 22.5k \\
Indomethacin & 5.2k & 3.1k & 598 & 8.9k \\
Lenalidomide & 1.3k & 601 & 56 & 1.9k \\
LY2624803 & 6.5k & 463 & 116 & 7.1k \\
LY2806920 & 4.7k & 660 & 59 & 5.4k \\
Mebendazole (tautomer 1) & 1.7k & 286 & 44 & 2.0k \\
Mebendazole (tautomer 2) & 2.4k & 254 & 45 & 2.7k \\
Mebendazole (tautomer 3) & 2.6k & 774 & 94 & 3.4k \\
Meloxicam & 2.2k & 521 & 55 & 2.8k \\
Mk-2022 & 23.0k & 5.4k & 350 & 28.8k \\
Mk-8876 & 19.1k & 2.9k & 600 & 22.6k \\
Olanzapine & 3.2k & 492 & 71 & 3.7k \\
Piracetam & 488 & 588 & 38 & 1.1k \\
Progesterone (enantiopure) & 4.0k & 1.5k & 199 & 5.7k \\
Progesterone (racemate) & 5.7k & 1.4k & 222 & 7.3k \\
Ripretinib & 22.9k & 1.9k & 136 & 25.0k \\
Rotigotine & 5.3k & 402 & 90 & 5.8k \\
ROY & 3.5k & 4.5k & 489 & 8.5k \\
Sildenafil & 19.3k & 3.0k & 383 & 22.7k \\
Sotagliflozin & 4.9k & 373 & 150 & 5.5k \\
Target XX & 19.3k & 7.1k & 537 & 26.9k \\
Target XXII & 1.2k & 790 & 44 & 2.0k \\
Target XXIII & 7.9k & 3.7k & 1.0k & 12.7k \\
TIk-301 & 3.1k & 1.8k & 138 & 5.1k \\
Tolbutamide & 6.4k & 6.2k & 669 & 13.3k \\
Uridine Triacetate & 10.5k & 526 & 77 & 11.1k \\
Vibegron & 3.8k & 29 & 43 & 3.9k \\
Voriconazole & 1.8k & 74 & 37 & 2.0k \\
\midrule
Deucravacitinib & 12.1k & 156 & 124 & 12.4k \\
Omaveloxone & 835 & 495 & 52 & 1.4k \\
Zuranolone & 6.8k & 595 & 76 & 7.4k \\
\midrule
\textbf{Average} & \textbf{6.4k} & \textbf{1.8k} & \textbf{209} & \textbf{8.4k} \\
\textbf{Median} & \textbf{3.9k} & \textbf{782} & \textbf{121} & \textbf{5.4k} \\
\textbf{Total} & \textbf{334.6k} & \textbf{92.4k} & \textbf{10.9k} & \textbf{437.9k} \\
\bottomrule
\end{tabular}
\end{table}

\subsubsection{Case Study: Rotigotine}

\begin{figure}[ht]
  \centering
  \includegraphics[width=8.8cm]{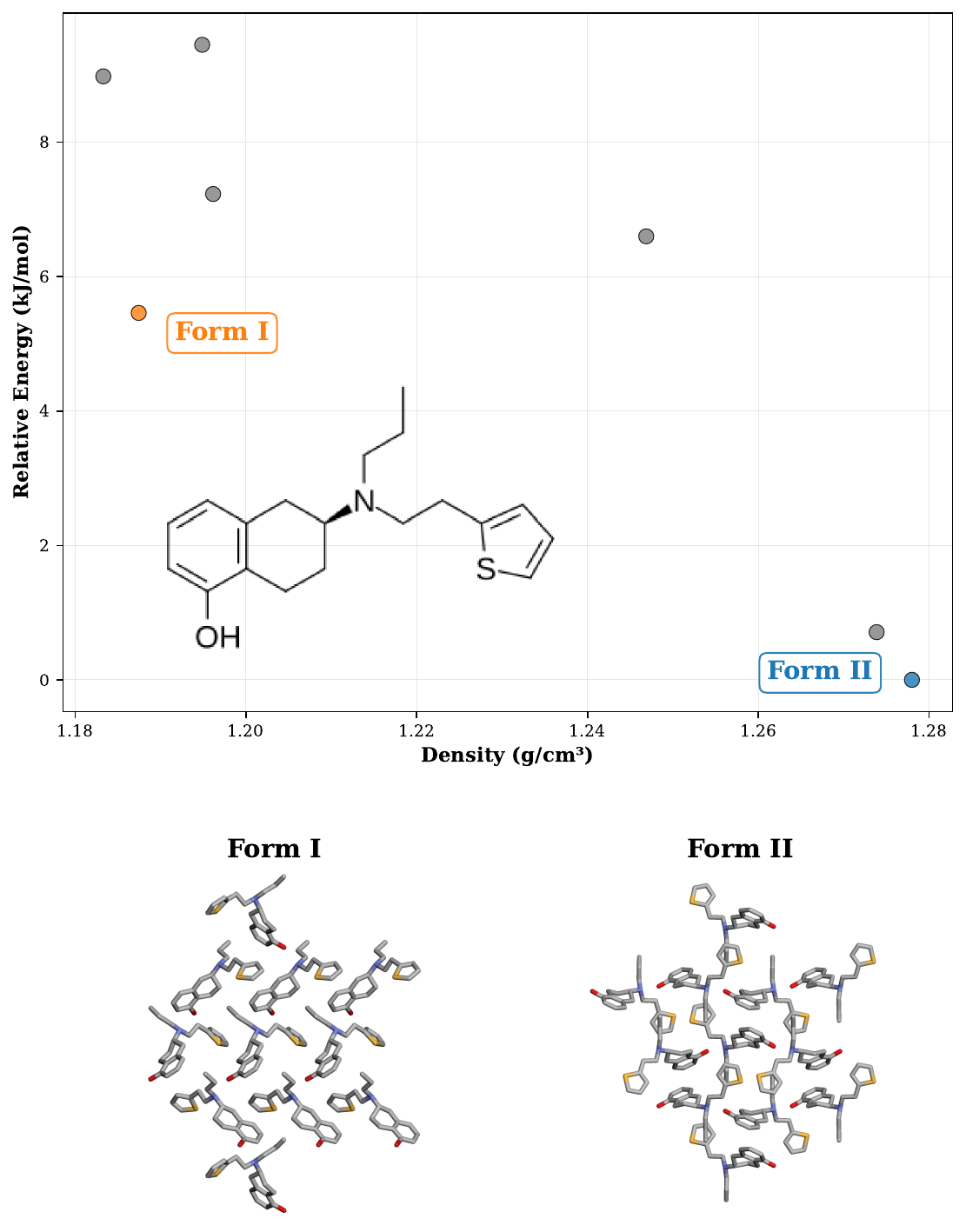}
  \caption{Density energy plot for Rotigotine. 
  Form I is the originally marketed form, but was later discovered to be metastable, with the more stable Form II discovered to be the thermodynamically stable form. 
  The Lavo CSP protocol generates and correctly ranks both forms. The 3D structures of Form I and II are shown, showing how molecules can pack into multiple polymorphs.}
  \label{fig:rotigotine}
\end{figure}

Rotigotine, a dopamine agonist used to treat Parkinson's disease, is a classic illustration of how CSP can mitigate the risk associated with late-appearing polymorphs. \cite{rotigotine, Frampton2019Rotigotine} 
Like ritonavir, the initially marketed form of rotigotine (Form I) was later found to be metastable, with a more stable polymorph (Form II) appearing unexpectedly, which led to product recalls and reformulation efforts. \cite{Rietveld2015, Chaudhuri01112008}
Application of the Lavo CSP protocol to rotigotine successfully identified and correctly ranked both forms. The crystal energy landscape, shown in Figure \ref{fig:rotigotine}, clearly identifies the experimentally observed Form II as the global energetic minimum. The metastable Form I was also found and predicted to be 5.45 kJ mol$^{-1}$ higher in energy than Form II. This computational result is in good agreement with the experimentally determined enthalpy difference of approximately 7.5 kJ mol$^{-1}$, validating the accuracy of the protocol. \cite{Mortazavi2019} If the developers had used this protocol, they would have known Form I was metastable and avoided this disaster.

\subsubsection{Case Study: Mebendazole}

\begin{figure}[ht]
  \centering
  \includegraphics[width=8.8cm]{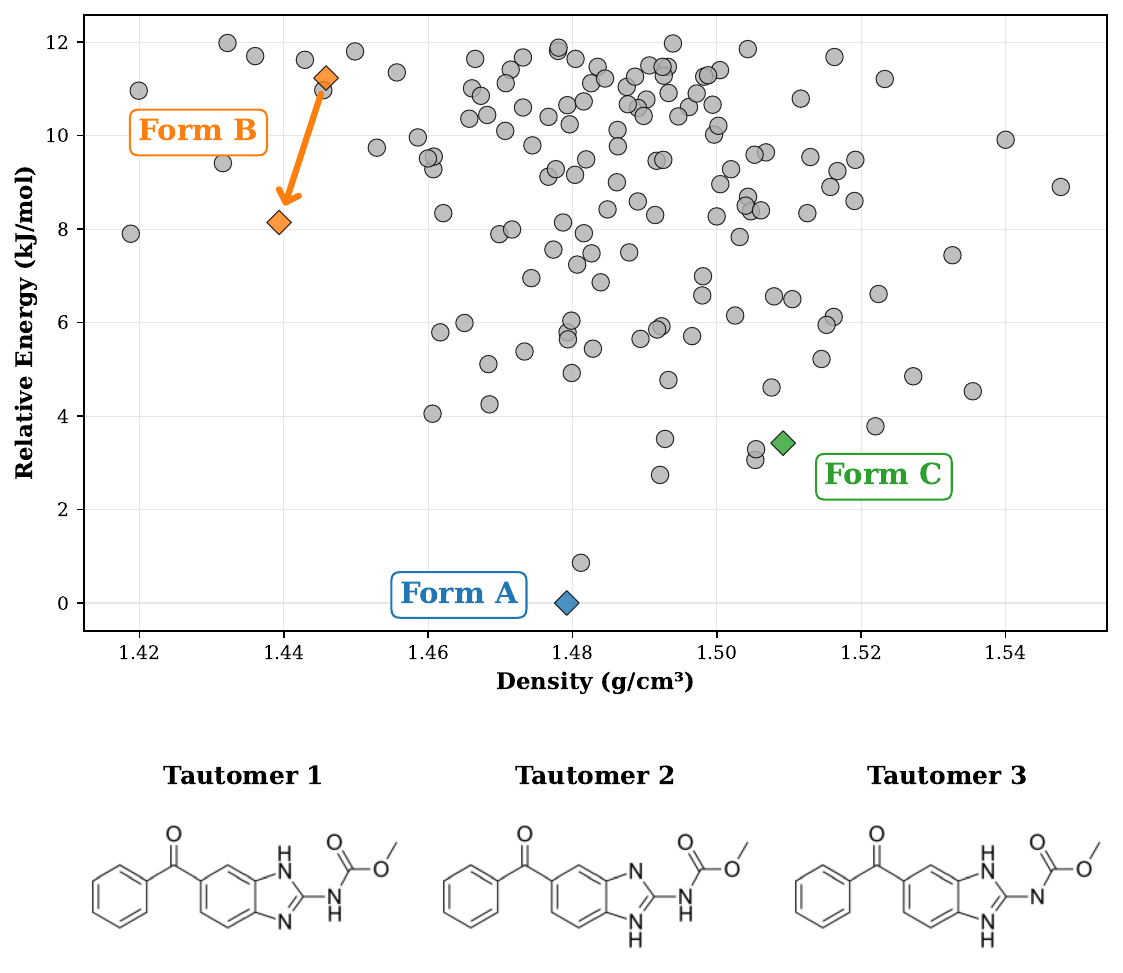}
  \caption{Density energy plot of the three polymorphs of mebendazole. The variant of Form B containing tautomer 2 is more stable, as indicated with the orange arrow. All three tautomers are found in experimental crystals.}
  \label{fig:mebendazole}
\end{figure}

The antihelmintic drug mebendazole presents an interesting case for crystal structure prediction due to the interplay of polymorphism and tautomerism. \cite{Bravetti2022Mebendazole, perry25} The benzimidazole-carbamate moiety of mebendazole can exist in three principal tautomeric forms (referred to as tautomers 1, 2, and 3), as illustrated in Figure \ref{fig:mebendazole}. Experimentally, three polymorphs are known, A, B, and C, which have been reported in the literature to contain tautomer 1, 3, and 2, respectively. The relative stability of these forms is of clinical importance; Form C is often preferred for therapeutic use as its intermediate stability offers a favorable balance between the poor bioavailability of Form A and the higher toxicity of Form B.

These CSP results corroborate recent findings which call into question the tautomer of the experimentally characterized form B structure. \cite{perry25} In the final CSP landscape, an exact match to the CSD entry for form B (containing tautomer 3) is found and ranked in the final crystal energy landscape at +11.2 kJ mol$^{-1}$). 
However, an analogous crystal structure  with the same packing but containing tautomer 2 was also generated and predicted to be significantly more stable (+8.2 kJ mol$^{-1}$). 
This result suggests that the true structure of form B contains tautomer 2. This conclusion aligns with the computational and solid-state NMR analysis presented by Perry et al. 
The CSP of mebendazole successfully identifies the known polymorphs and provides insights into their relative stability and tautomeric composition. The results of the study predict that form A is the global minimum structure. Forms C and B were found to be less stable, with relative lattice energies of +3.4 kJ mol$^{-1}$ and +8.2 kJ mol$^{-1}$, respectively. This predicted stability ranking (A $>$ C $>$ B) is consistent with experimental observations.
Furthermore, this case highlights the protocol's ability not only to predict the relative stability of known polymorphs but also to refine the structural understanding of complex pharmaceutical solids.

\subsubsection{Case Study: Fenofibrate}

\begin{figure}[!htbp]
  \includegraphics[width=\linewidth]{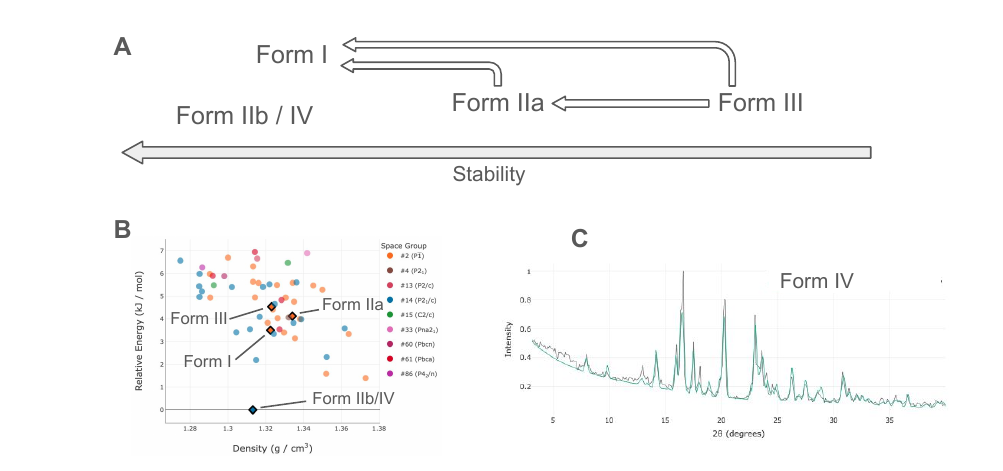}
  \caption{A qualitative summary of experimental forms of fenofibrate and their relative rankings, with white arrows indicating spontaneous interconversion at room temperature (A), the density-energy landscape from the CSP conducted in this work (B), and a refined theoretical powder pattern in green overlaid with the experimental Form IV powder pattern in gray (C).}
  \label{fig:fenofibrate}
\end{figure}

Fenofibrate is an oral medication used to treat abnormal blood lipid levels. \cite{Sidhu2023Fenofibrate} There are four entries in the CSD corresponding to the known neat, anhydrous solid forms of fenofibrate: Forms I, IIa, IIb, and III, according to the ontology used by Tipduangta et al. in 2018. \cite{C8RA01182F} Form I is a stable triclinic form generally used in formulation, where Form IIa is structurally very similar and is known to convert to Form I at room temperature. \cite{fenofibrate_IIa} Form IIb is a distinct, stable, monoclinic form isolated in 2012 and is believed to be most stable form. \cite{Balendiran2012_fenofibrate} Form III is likely the least stable, as it has a lower melting point, is difficult to isolate, and can easily convert to Forms IIa and I. Recently, a 2018 patent claims a stable, soluble Form IV for which there is no solved crystal structure, only a powder pattern, a DSC curve, and a FTIR spectrum. \cite{2018_fenofibrate_patent} Qualitative experimental stability rankings are summarized in Figure \ref{fig:fenofibrate}A.

All four forms with corresponding SCXRD data (I, IIa, IIb, III) are generated and identified as low-energy crystals on the theoretical landscape. We rank Form IIb as most stable, which agrees with the analysis of Balendiran, et al., followed by Forms I, IIa, and III. Ranking from CSP is shown in the density-energy landscape in Fig. \ref{fig:fenofibrate}B.
This relative ordering is consistent with experimental observations, which characterize Forms IIa and III as metastable conformational funnels that readily interconvert to Form I.

This CSP and subsequent analysis reveals interesting details with regards to the recent claim of a yet-undiscovered Form IV. In an effort to characterize the structure of Form IV, we refined simulated powder patterns of all generated structures to check if any matched the reported Form IV pattern. The best and only convincing powder pattern match, at almost 95\% similarity, is the crystal corresponding to Form IIb, overlay shown in Fig. \ref{fig:fenofibrate}C. 
The DSC shows a melting point similar to that of Form I, and is otherwise stable at room temperature, unlike Forms IIa and III. 
Due to the similarity in powder patterns and experimental observations of Forms IIb and IV, and the fact that the name Form II was ambiguous and overloaded until the clarification by Tipduangta et al. (after the 2018 patent), we believe it to be likely that Forms IIb and IV are in fact a single stable form. This study illustrates where computation can complement, and in this case correct, interpretation of experimental results.

\subsubsection{Case Study: Progesterone}

Progesterone, an endogenous steroid, naturally appears as a single enantiomer. When the starting material is known to be enantiopure, CSP is simply performed in the Sohncke space groups, those which exclude inversion centers. There are two known forms of enantiopure progesterone, both of which are sampled and ranked as low-energy in our Sohncke-only CSP. \cite{original_progesterone, progesterone_formII} Often, as is the case with progesterone, racemic mixtures can create denser, lower-energy crystals. As detailed in Ref. \cite{racemic_progesterone}, synthetic racemic progesterone crystallizes into a highly stable racemate. This work reproduces the findings of Price, \textit{et al.}, where CSP conducted in the non-Sohncke space groups predicts the known racemate to be the most stable among theoretical crystals, over 7 kJ mol$^{-1}$ lower in energy than the most stable enantiopure form. This is a scenario where computation guided experiment, suggesting the racemate must exist due to its stability relative to the enantiopure forms. 

\subsection{Blind PXRD Refinement} \label{ssec:pxrd_benchmark}
Crystalline powders of drug substances are relatively easy to prepare.
As a result, PXRD is a widely used method to obtain structural fingerprints of solid materials.
These fingerprints, or patterns, are typically sufficient to discriminate between the polymorphs of a drug molecule. \cite{Lee2014PolymorphGuide} However, powder patterns themselves are not sufficient to actually solve the 3-dimensional crystal structure.
By contrast, single crystal X-ray diffraction resolves the full crystal structure but requires slow growth of sufficiently large single crystals for imaging. Growing larger crystals is generally more time-consuming and often much more difficult or impossible using typical lab techniques. \cite{Newman2022Powders}
By simulating powder patterns of theoretical structures from a CSP screen, the theoretical structures can be compared directly to experimental powder patterns. \cite{OterodelaRoza2024PXRD}
In the common situation that only PXRD data is known for molecular crystals of a drug, CSP serves two purposes: 1. To quantify the stability of experimental polymorphs on the landscape relative to other possible polymorphs, and 2. To “solve” the unknown 3-dimensional atomic structure of the experimental polymorphs.

\begin{figure}[ht]
  \includegraphics[width=\textwidth]{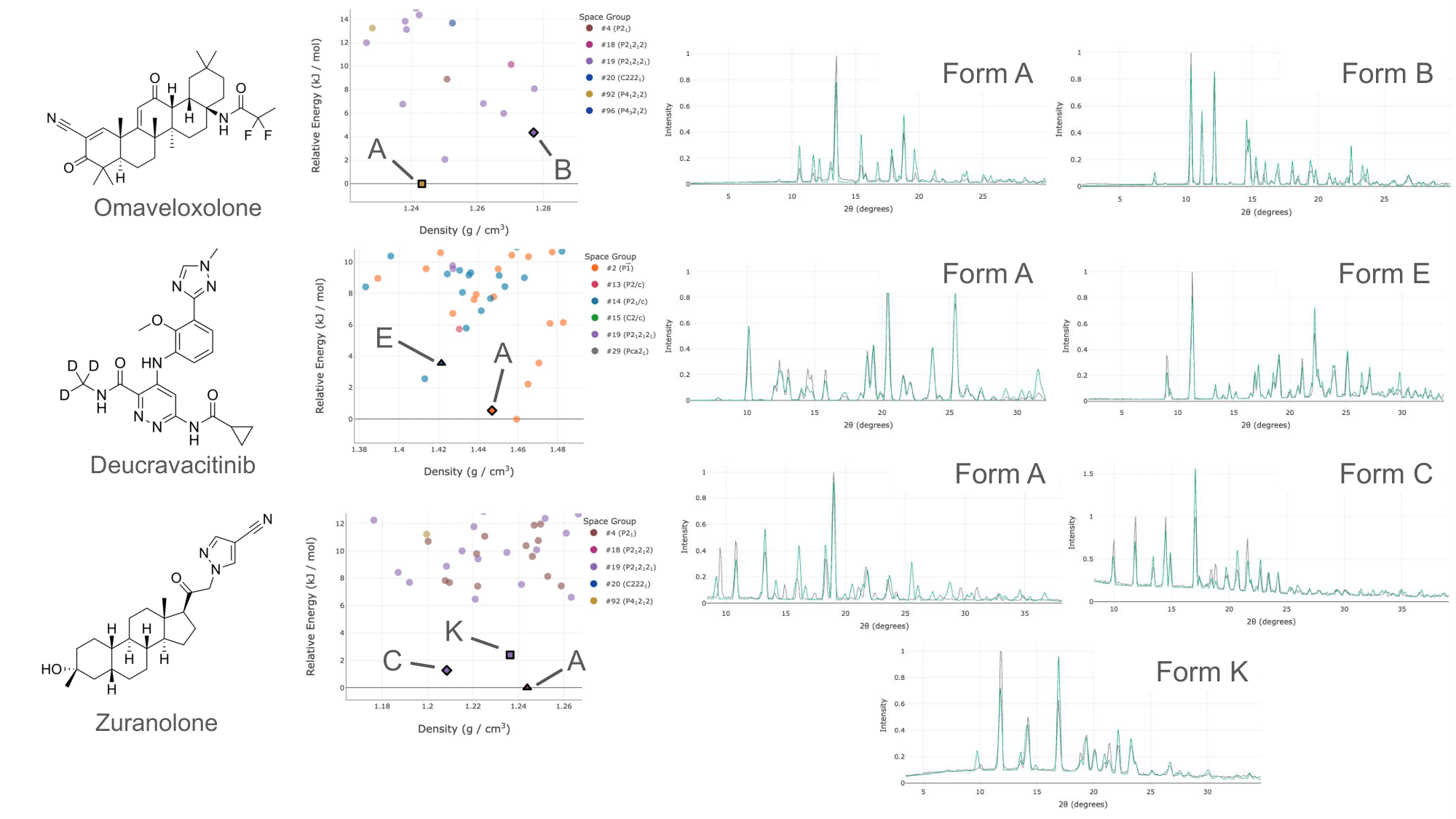}
  \caption{Summary of semi-blinded powder X-ray diffraction study. Rows correspond to each molecular structure (left). A low-energy subset of the crystal landscape from crystal structure prediction is provided for each molecule, with forms matching the experimental pattern indicated (center). Refined, simulated powder patterns of each experimental match are shown in green, overlaying the experimental pattern in gray (right).}
  \label{fig:pxrd}
\end{figure}

Three modern drugs were used to assess the ability of our crystal structure prediction protocol to assign structure and stability rankings to forms imaged only by PXRD: omaveloxolone, deucravacitinib, and zuranolone (Fig. \ref{fig:pxrd}). To our knowledge, there are no publicly-available SCXRD structures for these forms, only powder patterns reported in patents. \cite{omaveloxolone_forms, deucravacitinib_1, deucravacitinib_7, zuranolone_forms} Some references report unit cell parameters and space groups from internally-conducted SCXRD, but not molecular coordinates or conformations. We ran CSP on these molecules with the exact same protocol as the previous section, except the final crystal landscapes are compared to powder patterns extracted from patents and to reported unit cell parameters when available, rather than the atomic structural data contained in Crystallographic Information Files. For all forms in the resulting landscape, we simulate the powder pattern at the experimentally reported wavelength and refine the unit cell parameters and peak shape parameters to maximize a cross correlation-based similarity score. As in the previous section, we focus on crystalline, anhydrous forms. All three of these studies were conducted independently of patent authors as a semi-blinded proof-of-concept exercise. 

\subsubsection{Omaveloxolone}
Omaveloxolone is a first-in-class medication for the treatment of Friedreich’s ataxia and was approved for medical use in the United States in 2023. \cite{omaveloxolone_general} A 2020 patent on omaveloxolone describes two neat forms, Form A and Form B. \cite{omaveloxolone_forms} Form A is described as metastable, with a lower melting point than Form B and slight hygroscopicity indicated by mass loss in thermogravimetric analysis/mass spectrometry (TGA-MS). Despite its size, omaveloxolone is conformationally constrained due to its fused rings and thus required only 835 CPU hours to exhaustively sample its crystal landscape. The simulated patterns of two low-energy forms have notably high similarity scores when compared to the experimental Form A and Form B powder patterns, reproducing all reported peaks. The generated crystal that best matches the Form B pattern is the most stable on our final landscape, just 4.4 kJ mol$^{-1}$ more stable than the crystal that best matches the Form A pattern.

\subsubsection{Deucravacitinib}
Deucravacitinib is a first-in-class plaque psoriasis medication approved for medical use in the United States in 2022. \cite{deucravacitinib_general} The patent literature on the solid forms of deucravacitinib is prolific and growing: there are over 20 claimed crystalline forms of deucravacitinib across more than 9 unique patents, with additional patents related to amorphous dispersions. \cite{deucravacitinib_1, deucravacitinib_2, deucravacitinib_3, deucravacitinib_4, deucravacitinib_5, deucravacitinib_6, deucravacitinib_7, deucravacitinib_8, deucravacitinib_9} To constrain the search to forms covered by the CSP protocol used in this study, we only compared against the two forms validated by the patent authors with SCXRD, and for which we can ensure forms are neat and have only one molecule in the asymmetric unit. These are Forms A and E, declared in \cite{deucravacitinib_1} and \cite{deucravacitinib_7}, respectively. 
The 2018 patent disclosed crystalline Form A of this molecule, including a powder pattern, unit cell parameters, and the space group, but no resolved structure from SCXRD. 
The 2024 patent reports Form E, which the author’s internal SCXRD verifies is a free base, anhydrous form. 
Differential scanning calorimetry (DSC) measures a lower transition temperature for Form E than Form A, suggesting Form E is marginally less thermodynamically stable than the originally patented Form A. 

Deucravacitinib has significant conformational flexibility and it required 12.1k CPU hours to exhaustively sample its crystal landscape. 
The resulting DFT-ranked landscape contained theoretical forms that substantially reproduced the patterns corresponding to Forms A and E. 
The theoretical form that best reproduces the Form A pattern is in the reported space group P$\bar{1}$, and its unit cell parameters match to within 4\%. 
This theoretical form is the second most stable form on the landscape, with a lattice energy just 0.55 kJ mol$^{-1}$ higher than the predicted lowest-energy form. 
The theoretical form that best matches the Form E pattern is in the experimental space group P2$_1$/n and matches all unit cell parameters to within 3\%. 
This form lies 3.01 kJ mol$^{-1}$ higher in energy than the Form A match, qualitatively agreeing with the stability ranking suggested by DSC.

\subsubsection{Zuranolone}
Zuranolone is an orally-administrated medication used to treat postpartum depression, approved for medical use in the United States in late 2023. \cite{zuranolone_general} A 2018 patent describes three anhydrous crystalline forms in addition to a number of solvates and metastable forms. \cite{zuranolone_forms} In the present study, we attempted to identify the three known anhydrous forms known as Forms A, C, and K. The authors did grow single crystals and perform SCXRD on Forms A and C, which they assert are more thermodynamically favorable. However, they reported only unit cell parameters of each form and single images from a molecular visualization software, omitting specific structural information. The authors indicate that Form K is a high-temperature form, as many other crystalline forms interconvert to Form K at elevated temperature, most notably the interconversion of Form A crystals to Form K above 157.2 \textdegree C. The authors report only a PXRD pattern of Form K. All three forms have closely matching powder patterns on the CSP landscape, and the unit cell parameters of the structures corresponding to powder patterns of A and C all differ by less than 3\%. The three matching theoretical forms are the lowest-energy forms on the theoretical landscape, with stability qualitatively matching experimental observations: Form A $>$ Form C (+1.26 kJ mol$^{-1}$) $>$ Form K (+2.40 kJ mol$^{-1}$).

\subsection{Comparison to other NNPs}

\begin{figure}[ht]
  \centering
  \includegraphics[]{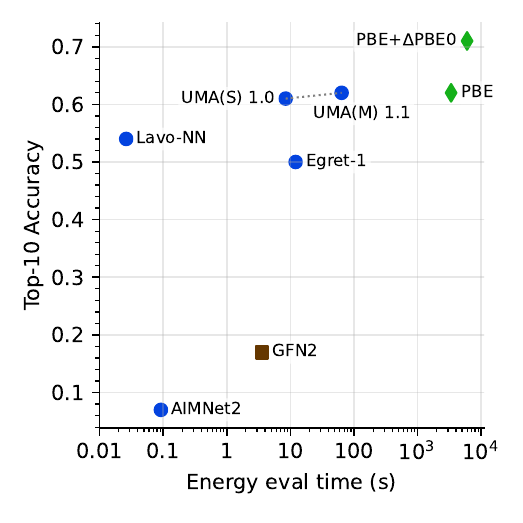}
  \caption{Cost-accuracy comparison of various NNPs (blue circles), semi-empirical methods (brown squares), and DFT functionals (green diamonds) on a large CSP ranking task.}
  \label{fig:nnp-pareto}
\end{figure}

To contextualize the performance of Lavo-NN, the crystal structure landscapes generated and described in Section \ref{ssec:benchmark} were used as a benchmark to assess the ranking ability and efficiency of various NNPs and other quantum chemistry methods.
The practical value of a potential is arguably best related to how reliably the potential ranks experimental polymorphs near the bottom of a landscape, rather than correlation in energies between the potential and some high-level reference method.
Therefore, the accuracy of a potential was quantified by the number of experimental polymorphs which appear in the bottom ten structures of their respective landscapes, referred to as ``Top-10 Accuracy".
Similar metrics are used in assessing docking methods for drug discovery.
The cost of each potential is defined as the single-core CPU wall time required to perform an energy evaluation, geometrically averaged over all 104 polymorphs.
Although single-core CPU calculations are not necessarily representative of average hardware, this measure allows for straightforward relative comparisons between potentials.
These cost and accuracy results are displayed in Figure \ref{fig:nnp-pareto}, which contains a pareto plot of the various potentials.
The DFT re-ranking methods discussed in Section \ref{ssec:dft_reranking}, PBE and PBE+
$\Delta$PBE0, are included.
These methods are highly reliable but slow, so they serve as a reference against which the speed and accuracy of the NNPs can be judged.

It is clear from this figure that Lavo-NN occupies a highly favorable position on the cost-accuracy frontier. 
Lavo-NN achieves a Top-10 accuracy of 54\% while being almost three orders of magnitude faster than the next-fastest model of equal or better accuracy, UMA small v1.0 (OMC head), which scores 61 \%. \cite{uma}
These two models, Lavo-NN and UMA small v1.0, form a pareto front relative to the other NNPs.
It is interesting to note that the medium-sized UMA model, which is trained on the same data as the smaller UMA model, fares no better at this CSP ranking task while being much more expensive.
This could suggest that the current limitation of the UMA models here is not model capacity (e.g. number of parameters), but either the expressivity of the model or the quality and/or breadth of the training dataset.
Given that UMA is trained on approximately 500M DFT data points, some of which are high-accuracy periodic molecular crystal structures, it is most likely that improvements to the underlying architecture would be the best way to improve this already impressive model.

The Egret-1 model, a carefully trained instance of the MACE architecture, serves as an interesting contrast to both Lavo-NN and the UMA models. \cite{egret1,mace-off}
Egret-1 is similar in cost to the smaller of the two UMA models, but is less accurate (scoring 50\% in the Top-10 metric), no doubt in part due to the large disparity in training data; Egret-1 was trained on only $\sim$ 1M points from the SPICE dataset.
This same data was a significant part of Lavo-NN's training set, and the two models (Lavo-NN and Egret-1) exhibit similar accuracy, but differ in cost by almost three orders of magnitude.
Lastly, the general purpose AIMNet2 model is the fastest NNP apart from Lavo-NN, but it struggles in the ranking task, with a Top-10 accuracy of under 10\% without system-specific fine-tuning. \cite{Anstine2025AIMNet2}
The widely used semi-empirical GFN2-xTB method, often considered a robust alternative to NNPs given its similar speed, proves to be less accurate than most of the NNPs. \cite{GFN2}.

Lavo-NN's unique combination of high speed and robust accuracy makes it exceptionally well-suited for the primary generation phase of CSP, where hundreds of millions of energy evaluations are required, and more expensive models like UMA and Egret-1 are not tractable. 
This result demonstrates that a domain-specific architecture and a targeted training dataset can produce a potential that is not only fast enough for large-scale crystal structure generation but is also accurate enough to reliably identify experimentally relevant polymorphs.

\section{Conclusions and Outlook}
This study introduces and validates a protocol for automated, low-cost crystal structure prediction of small molecule drugs.
Central to the approach is Lavo-NN, a domain-specific NNP that both generates and accurately ranks crystal polymorphs at multiple orders of magnitude higher throughput than the next-best publicly benchmarked NNP, UMA small v1.0, while demonstrating accuracy close to periodic DFT. 
A cloud-parallel implementation enabled a benchmark study of 49 molecules (47 of which are drugs) and 110 experimental polymorphs (characterized by either SCXRD or PXRD), making this the largest pharmaceutical CSP benchmark to date.
The total cost of all CSP tasks was under 438k CPU hours, which is economical considering the size and complexity of many of the 49 molecules included.
Several case studies from this benchmark illustrate the role of efficient CSP in complementing and clarifying experimental efforts.
The CSP protocol discussed here has the potential to help drug developers de-risk solid form more quickly and routinely.
Falling compute costs should open routine CSP to larger or earlier-stage compounds.

While advancements in CSP are exciting, many challenges remain.
A top priority is extending the protocol described here to crystals containing multiple independent molecules in the asymmetric unit ($Z' > 1$), not to mention hydrates, salts, and co-crystals.
Crystals such as these have more degrees of freedom than the $Z' = 1$ crystals benchmarked here, making generation challenging because the search space grows at worst exponentially with these degrees of freedom.
In contrast, the focus of this work was reducing the bottleneck of the downstream \textit{ranking} task.
For 40 of 49 total molecules in this study, the cost of generating a landscape of low-energy crystal structures exceeded the cost of re-ranking the landscape with single-point DFT energies. 
Put another way, as ranking costs decrease with improving potentials, generation will become the dominant cost of CSP.
Affordable multi-component crystal structure generation may therefore require new insights, similar to the generative models that have reshaped the protein sequence-to-structure prediction problem. \cite{jumper2021highly}
Moreover, CSP is a de-risking activity: failing to generate even one experimentally accessible form is a disqualifying failure, unlike protein folding where approximate models can still be useful.
Long term, computational methods seek to model not just thermodynamically favorable crystal structures, but to connect these crystal structures to experimental conditions.
This includes predicting stability with respect to temperature, pressure, and relative humidity as well as understanding the kinetic processes which govern crystallization and lead to the formation of metastable polymorphs.
Solutions to the problems above work in service of a still-distant final frontier: reliable prediction of all polymorphic forms, and the experimental procedure to grow each in the lab.

Looking forward, we expect NNPs to represent a growing part of typical CSP protocols, replacing expensive DFT computations. 
This is only now feasible for real-world pharmaceutical molecules, as previous pre-trained NNPs lack either the speed or the accuracy for this particular domain.
Beyond crystal structure prediction, combining physics-based models with task-oriented data and hardware-aware architectures enables NNPs to achieve the robustness and efficiency which surpasses general models.

\appendix % if required
\section{Appendix}

\subsection{Data Availability}
All crystal structure landscapes generated in the CSP benchmark, in the form of Crystallographic Information Files (CIFs) and associated energies, will be made publicly available in a GitHub repository upon publication of this article.

% \begin{acknowledgements}

% \end{acknowledgements}

\begin{funding}
This material is based upon work supported by the National Science Foundation
SBIR Phase I Award No.\ 2227936.%
\newline
Any opinions, findings, and conclusions or recommendations expressed in this
material are those of the authors and do not necessarily reflect the views of
the National Science Foundation.
\end{funding}

\ConflictsOfInterest{The authors declare no competing interests.
}

\bibliography{iucr} % basename of .bib file

@article{perry25, 
    title={Taming Tautomerism in Organic Crystal Structure Prediction}, 
    url={https://doi.org/10.26434/chemrxiv-2025-288td}, 
    DOI={10.26434/chemrxiv-2025-288td}, 
    journal={ChemRxiv}, 
    author={Perry, Cody and Ramos, Sebastian and Phelps, Maria and Mueller, Leonard and Beran, Gregory}, 
    year={2025}, 
    month=may, 
    language={en} }

@article{Pudipeddi2005,
  title = {Trends in Solubility of Polymorphs},
  volume = {94},
  ISSN = {0022-3549},
  url = {http://dx.doi.org/10.1002/jps.20302},
  DOI = {10.1002/jps.20302},
  number = {5},
  journal = {Journal of Pharmaceutical Sciences},
  publisher = {Elsevier BV},
  author = {Pudipeddi,  Madhu and Serajuddin,  Abu T.M.},
  year = {2005},
  month = may,
  pages = {929–939}
}

@article{Listiohadi2008,
  title = {Moisture sorption,  compressibility and caking of lactose polymorphs},
  volume = {359},
  ISSN = {0378-5173},
  url = {http://dx.doi.org/10.1016/j.ijpharm.2008.03.044},
  DOI = {10.1016/j.ijpharm.2008.03.044},
  number = {1–2},
  journal = {International Journal of Pharmaceutics},
  publisher = {Elsevier BV},
  author = {Listiohadi,  Y. and Hourigan,  J.A. and Sleigh,  R.W. and Steele,  R.J.},
  year = {2008},
  month = jul,
  pages = {123–134}
}

@Article{C5CS00227C,
author ="Cruz-Cabeza, Aurora J. and Reutzel-Edens, Susan M. and Bernstein, Joel",
title  ="Facts and fictions about polymorphism",
journal  ="Chem. Soc. Rev.",
year  ="2015",
volume  ="44",
issue  ="23",
pages  ="8619-8635",
publisher  ="The Royal Society of Chemistry",
doi  ="10.1039/C5CS00227C",
url  ="http://dx.doi.org/10.1039/C5CS00227C",
}

@article{Lin2014,
  title = {An Overview of Famotidine Polymorphs: Solid-State Characteristics,  Thermodynamics,  Polymorphic Transformation and Quality Control},
  volume = {31},
  ISSN = {1573-904X},
  url = {http://dx.doi.org/10.1007/s11095-014-1323-5},
  DOI = {10.1007/s11095-014-1323-5},
  number = {7},
  journal = {Pharmaceutical Research},
  publisher = {Springer Science and Business Media LLC},
  author = {Lin,  Shan-Yang},
  year = {2014},
  month = mar,
  pages = {1619–1631}
}

@article{EGART2014347,
title = {Compaction properties of crystalline pharmaceutical ingredients according to the Walker model and nanomechanical attributes},
journal = {International Journal of Pharmaceutics},
volume = {472},
number = {1},
pages = {347-355},
year = {2014},
issn = {0378-5173},
doi = {https://doi.org/10.1016/j.ijpharm.2014.06.047},
url = {https://www.sciencedirect.com/science/article/pii/S0378517314004761},
author = {M. Egart and I. Ilić and B. Janković and N. Lah and S. Srčič},
}

@article{Bauer2001,
  volume = {18},
  ISSN = {0724-8741},
  url = {http://dx.doi.org/10.1023/a:1011052932607},
  DOI = {10.1023/a:1011052932607},
  number = {6},
  journal = {Pharmaceutical Research},
  publisher = {Springer Science and Business Media LLC},
  author = {Bauer,  John and Spanton,  Stephen and Henry,  Rodger and Quick,  John and Dziki,  Walter and Porter,  William and Morris,  John},
  year = {2001},
  pages = {859–866}
}

@misc{Elder2017,
  title = {ICH Q6A Specifications: Test Procedures and Acceptance Criteria for New Drug Substances and New Drug Products: Chemical Substances},
  ISBN = {9781118971147},
  url = {http://dx.doi.org/10.1002/9781118971147.ch16},
  DOI = {10.1002/9781118971147.ch16},
  journal = {ICH Quality Guidelines},
  publisher = {Wiley},
  author = {Elder,  David},
  year = {2017},
  month = sep,
  pages = {433–466}
}

@misc{USFDA2010,
  title = {{Guidance for Industry on Drug Substance Chemistry, Manufacturing, and Controls Information}},
  author = {{U.S. Food and Drug Administration}},
  year = {2010},
  month = aug,
  howpublished = {Federal Register},
  url = {https://www.federalregister.gov/documents/2010/08/06/2010-19360/guidance-for-industry-on-drug-substance-chemistry-manufacturing-and-controls-information},
}

@article{kattan2015what,
  title={What Form Is It? Lessons from 13 Polymorph Pharmaceutical Cases},
  author={Sodikoff, Brian and Masar, Martin and Garrett, Cole},
  journal={Katten Muchin Rosenman LLP},
  year={2015},
  url="{Link: Katten Muchin Rosenman LLP https://katten.com/what-form-is-it-lessons-from-13-polymorph-pharmaceutical-cases}"
}

@article{Groom2016,
  author = {Groom, C. R. and Bruno, I. J. and Lightfoot, M. P. and Ward, S. C.},
  title = {The Cambridge Structural Database},
  journal = {Acta Crystallographica, Section B: Structural Science, Crystal Engineering and Materials},
  volume = {72},
  issue = {2},
  pages = {171-179},
  year = {2016},
  doi = {10.1107/S2052520616003954},
}

@Article{C8RA01182F,
author ="Tipduangta, Pratchaya and Takieddin, Khaled and Fábián, László and Belton, Peter and Qi, Sheng",
title  ="Towards controlling the crystallisation behaviour of fenofibrate melt: triggers of crystallisation and polymorphic transformation",
journal  ="RSC Adv.",
year  ="2018",
volume  ="8",
issue  ="24",
pages  ="13513-13525",
publisher  ="The Royal Society of Chemistry",
doi  ="10.1039/C8RA01182F",
url  ="http://dx.doi.org/10.1039/C8RA01182F",
}

@article{Eastman2023,
  title = {SPICE,  A Dataset of Drug-like Molecules and Peptides for Training Machine Learning Potentials},
  volume = {10},
  ISSN = {2052-4463},
  url = {http://dx.doi.org/10.1038/s41597-022-01882-6},
  DOI = {10.1038/s41597-022-01882-6},
  number = {1},
  journal = {Scientific Data},
  publisher = {Springer Science and Business Media LLC},
  author = {Eastman,  Peter and Behara,  Pavan Kumar and Dotson,  David L. and Galvelis,  Raimondas and Herr,  John E. and Horton,  Josh T. and Mao,  Yuezhi and Chodera,  John D. and Pritchard,  Benjamin P. and Wang,  Yuanqing and De Fabritiis,  Gianni and Markland,  Thomas E.},
  year = {2023},
  month = jan 
}

@article{Spronk2023,
  title = {A quantum chemical interaction energy dataset for accurately modeling protein-ligand interactions},
  volume = {10},
  ISSN = {2052-4463},
  url = {http://dx.doi.org/10.1038/s41597-023-02443-1},
  DOI = {10.1038/s41597-023-02443-1},
  number = {1},
  journal = {Scientific Data},
  publisher = {Springer Science and Business Media LLC},
  author = {Spronk,  Steven A. and Glick,  Zachary L. and Metcalf,  Derek P. and Sherrill,  C. David and Cheney,  Daniel L.},
  year = {2023},
  month = sep 
}

@misc{aws,
  author = {{Amazon Web Services}},
  title = {{Amazon Web Services}},
  url = {https://aws.amazon.com/},
  year = {2023}
}

@misc{aws-s3,
  author = {{Amazon Web Services}},
  title = {{Amazon S3}},
  url = {https://aws.amazon.com/s3},
  year = {2023}
}

@misc{aws-sqs,
  author = {{Amazon Web Services}},
  title = {{Amazon SQS}},
  url = {https://aws.amazon.com/sqs},
  year = {2023}
}

@misc{aws-batch,
  author = {{Amazon Web Services}},
  title = {{AWS Batch}},
  url = {https://aws.amazon.com/batch},
  year = {2023}
}

@misc{sqlite2025hipp, 
  title={{SQLite}},
  url={https://www.sqlite.org/index.html},
  version={3.50.2},
  year={2025},
  author={Hipp, Richard D}
}

@article{docker,
author = {Boettiger, Carl},
title = {An introduction to Docker for reproducible research},
year = {2015},
issue_date = {January 2015},
publisher = {Association for Computing Machinery},
address = {New York, NY, USA},
volume = {49},
number = {1},
issn = {0163-5980},
url = {https://doi.org/10.1145/2723872.2723882},
doi = {10.1145/2723872.2723882},
month = jan,
pages = {71–79},
numpages = {9}
}

@article{LIN2022103718,
title = {Methods for improving the availability of spot instances: A survey},
journal = {Computers in Industry},
volume = {141},
pages = {103718},
year = {2022},
issn = {0166-3615},
doi = {https://doi.org/10.1016/j.compind.2022.103718},
url = {https://www.sciencedirect.com/science/article/pii/S0166361522001154},
author = {Liduo Lin and Li Pan and Shijun Liu},
}

@book{heath1986hypercube,
  title={Hypercube multiprocessors 1986},
  author={Heath, Michael T and others},
  year={1986},
  publisher={Siam}
}

@book{foster1995designing,
  title={Designing and building parallel programs: concepts and tools for parallel software engineering},
  author={Foster, Ian},
  year={1995},
  publisher={Addison-Wesley Longman Publishing Co., Inc.}
}

@book{polymorphism_pharma_solids_brittain,
  title = {Polymorphism in Pharmaceutical Solids (2nd ed.)},
  author = {Harry G. Brittain},
  ISBN = {9780429147661},
  url = {http://dx.doi.org/10.3109/9781420073225},
  DOI = {10.3109/9781420073225},
  publisher = {CRC Press},
  year = {2009},
}

@article{Yao2023,
  title = {Ritonavir Form III: A New Polymorph After 24 Years},
  volume = {112},
  ISSN = {0022-3549},
  url = {http://dx.doi.org/10.1016/j.xphs.2022.09.026},
  DOI = {10.1016/j.xphs.2022.09.026},
  number = {1},
  journal = {Journal of Pharmaceutical Sciences},
  publisher = {Elsevier BV},
  author = {Yao,  Xin and Henry,  Rodger F. and Zhang,  Geoff G.Z.},
  year = {2023},
  month = jan,
  pages = {237–242}
}

@article{Catlow2023,
  title = {Crystal structure prediction: achievements and opportunities},
  volume = {10},
  ISSN = {2052-2525},
  url = {http://dx.doi.org/10.1107/S2052252523001835},
  DOI = {10.1107/s2052252523001835},
  number = {2},
  journal = {IUCrJ},
  publisher = {International Union of Crystallography (IUCr)},
  author = {Catlow,  C. Richard A.},
  year = {2023},
  month = mar,
  pages = {143–144}
}

@article{lommerse2000first,
  title={A test of crystal structure prediction of small organic molecules},
  author={Lommerse, J P M and Motherwell, W D S and Ammon, H L and Dunitz, J D and Gavezzotti, A and Hofmann, D W M and Leusen, F J J and Mooij, W T M and Price, S L and Schweizer, B and others},
  journal={Acta Crystallographica Section B: Structural Science},
  volume={56},
  number={4},
  pages={697--714},
  year={2000},
  publisher={International Union of Crystallography}
}

@article{motherwell2002second,
  title={Crystal structure prediction of small organic molecules: a second blind test},
  author={Motherwell, W D S and Ammon, H L and Dunitz, J D and Dzyabchenko, A and Erk, P and Gavezzotti, A and Hofmann, D W M and Leusen, F J J and Lommerse, J P M and Mooij, W T M and others},
  journal={Acta Crystallographica Section B: Structural Science},
  volume={58},
  number={4},
  pages={647--661},
  year={2002},
  publisher={International Union of Crystallography}
}

@article{day2005third,
  title={A third blind test of crystal-structure prediction},
  author={Day, Graeme M and Motherwell, WDS and Ammon, Herman L and Boerrigter, Stefan XM and Della Valle, RG and Dzyabchenko, A and Erk, P and Gavezzotti, A and Hofmann, DWM and Leusen, FJJ and others},
  journal={Acta Crystallographica Section B: Structural Science},
  volume={61},
  number={5},
  pages={511--527},
  year={2005},
  publisher={International Union of Crystallography}
}

@article{day2009fourth,
  title={Significant progress in predicting the crystal structures of small organic molecules--a report on the fourth blind test},
  author={Day, Graeme M and Cooper, T and Cruz-Cabeza, AJ and Hejczyk, KE and Ammon, HL and Boerrigter, SXM and Tan, JS and Della Valle, RG and Venuti, E and Jose, J and others},
  journal={Acta Crystallographica Section B: Structural Science},
  volume={65},
  number={2},
  pages={107--125},
  year={2009},
  publisher={International Union of Crystallography}
}

@article{bardwell2011fifth,
  title={Towards crystal structure prediction of complex organic compounds--a report on the fifth blind test},
  author={Bardwell, David A and Collins, Colin S and Hukin, Claire L and Labet, Vincent and Lunt, Geoffrey R and Mohammed, Raoof and Neumann, Marcus A and Price, Sarah L and Schmidt, Martin U and Day, Graeme M},
  journal={Acta Crystallographica Section B: Structural Science},
  volume={67},
  number={6},
  pages={535--551},
  year={2011},
  publisher={International Union of Crystallography}
}

@article{Reilly2016,
  title = {Report on the sixth blind test of organic crystal structure prediction methods},
  volume = {72},
  ISSN = {2052-5206},
  url = {http://dx.doi.org/10.1107/S2052520616007447},
  DOI = {10.1107/s2052520616007447},
  number = {4},
  journal = {Acta Crystallographica Section B Structural Science,  Crystal Engineering and Materials},
  publisher = {International Union of Crystallography (IUCr)},
  author = {Reilly,  Anthony M. and Cooper,  Richard I. and Adjiman,  Claire S. and Bhattacharya,  Saswata and Boese,  A. Daniel and Brandenburg,  Jan Gerit and Bygrave,  Peter J. and Bylsma,  Rita and Campbell,  Josh E. and Car,  Roberto and Case,  David H. and Chadha,  Renu and Cole,  Jason C. and Cosburn,  Katherine and Cuppen,  Herma M. and Curtis,  Farren and Day,  Graeme M. and DiStasio Jr,  Robert A. and Dzyabchenko,  Alexander and van Eijck,  Bouke P. and Elking,  Dennis M. and van den Ende,  Joost A. and Facelli,  Julio C. and Ferraro,  Marta B. and Fusti-Molnar,  Laszlo and Gatsiou,  Christina-Anna and Gee,  Thomas S. and de Gelder,  René and Ghiringhelli,  Luca M. and Goto,  Hitoshi and Grimme,  Stefan and Guo,  Rui and Hofmann,  Detlef W. M. and Hoja,  Johannes and Hylton,  Rebecca K. and Iuzzolino,  Luca and Jankiewicz,  Wojciech and de Jong,  Daniël T. and Kendrick,  John and de Klerk,  Niek J. J. and Ko,  Hsin-Yu and Kuleshova,  Liudmila N. and Li,  Xiayue and Lohani,  Sanjaya and Leusen,  Frank J. J. and Lund,  Albert M. and Lv,  Jian and Ma,  Yanming and Marom,  Noa and Masunov,  Artëm E. and McCabe,  Patrick and McMahon,  David P. and Meekes,  Hugo and Metz,  Michael P. and Misquitta,  Alston J. and Mohamed,  Sharmarke and Monserrat,  Bartomeu and Needs,  Richard J. and Neumann,  Marcus A. and Nyman,  Jonas and Obata,  Shigeaki and Oberhofer,  Harald and Oganov,  Artem R. and Orendt,  Anita M. and Pagola,  Gabriel I. and Pantelides,  Constantinos C. and Pickard,  Chris J. and Podeszwa,  Rafal and Price,  Louise S. and Price,  Sarah L. and Pulido,  Angeles and Read,  Murray G. and Reuter,  Karsten and Schneider,  Elia and Schober,  Christoph and Shields,  Gregory P. and Singh,  Pawanpreet and Sugden,  Isaac J. and Szalewicz,  Krzysztof and Taylor,  Christopher R. and Tkatchenko,  Alexandre and Tuckerman,  Mark E. and Vacarro,  Francesca and Vasileiadis,  Manolis and Vazquez-Mayagoitia,  Alvaro and Vogt,  Leslie and Wang,  Yanchao and Watson,  Rona E. and de Wijs,  Gilles A. and Yang,  Jack and Zhu,  Qiang and Groom,  Colin R.},
  year = {2016},
  month = aug,
  pages = {439–459}
}

@article{Hunnisett2024,
  title = {The seventh blind test of crystal structure prediction: structure generation methods},
  volume = {80},
  ISSN = {2052-5206},
  url = {http://dx.doi.org/10.1107/S2052520624007492},
  DOI = {10.1107/s2052520624007492},
  number = {6},
  journal = {Acta Crystallographica Section B Structural Science,  Crystal Engineering and Materials},
  publisher = {International Union of Crystallography (IUCr)},
  author = {Hunnisett,  Lily M. and Nyman,  Jonas and Francia,  Nicholas and Abraham,  Nathan S. and Adjiman,  Claire S. and Aitipamula,  Srinivasulu and Alkhidir,  Tamador and Almehairbi,  Mubarak and Anelli,  Andrea and Anstine,  Dylan M. and Anthony,  John E. and Arnold,  Joseph E. and Bahrami,  Faezeh and Bellucci,  Michael A. and Bhardwaj,  Rajni M. and Bier,  Imanuel and Bis,  Joanna A. and Boese,  A. Daniel and Bowskill,  David H. and Bramley,  James and Brandenburg,  Jan Gerit and Braun,  Doris E. and Butler,  Patrick W. V. and Cadden,  Joseph and Carino,  Stephen and Chan,  Eric J. and Chang,  Chao and Cheng,  Bingqing and Clarke,  Sarah M. and Coles,  Simon J. and Cooper,  Richard I. and Couch,  Ricky and Cuadrado,  Ramon and Darden,  Tom and Day,  Graeme M. and Dietrich,  Hanno and Ding,  Yiming and DiPasquale,  Antonio and Dhokale,  Bhausaheb and van Eijck,  Bouke P. and Elsegood,  Mark R. J. and Firaha,  Dzmitry and Fu,  Wenbo and Fukuzawa,  Kaori and Glover,  Joseph and Goto,  Hitoshi and Greenwell,  Chandler and Guo,  Rui and Harter,  J\"{u}rgen and Helfferich,  Julian and Hofmann,  Detlef W. M. and Hoja,  Johannes and Hone,  John and Hong,  Richard and Hutchison,  Geoffrey and Ikabata,  Yasuhiro and Isayev,  Olexandr and Ishaque,  Ommair and Jain,  Varsha and Jin,  Yingdi and Jing,  Aling and Johnson,  Erin R. and Jones,  Ian and Jose,  K. V. Jovan and Kabova,  Elena A. and Keates,  Adam and Kelly,  Paul F. and Khakimov,  Dmitry and Konstantinopoulos,  Stefanos and Kuleshova,  Liudmila N. and Li,  He and Lin,  Xiaolu and List,  Alexander and Liu,  Congcong and Liu,  Yifei Michelle and Liu,  Zenghui and Liu,  Zhi-Pan and Lubach,  Joseph W. and Marom,  Noa and Maryewski,  Alexander A. and Matsui,  Hiroyuki and Mattei,  Alessandra and Mayo,  R. Alex and Melkumov,  John W. and Mohamed,  Sharmarke and Momenzadeh Abardeh,  Zahrasadat and Muddana,  Hari S. and Nakayama,  Naofumi and Nayal,  Kamal Singh and Neumann,  Marcus A. and Nikhar,  Rahul and Obata,  Shigeaki and O’Connor,  Dana and Oganov,  Artem R. and Okuwaki,  Koji and Otero-de-la-Roza,  Alberto and Pantelides,  Constantinos C. and Parkin,  Sean and Pickard,  Chris J. and Pilia,  Luca and Pivina,  Tatyana and Podeszwa,  Rafał and Price,  Alastair J. A. and Price,  Louise S. and Price,  Sarah L. and Probert,  Michael R. and Pulido,  Angeles and Ramteke,  Gunjan Rajendra and Rehman,  Atta Ur and Reutzel-Edens,  Susan M. and Rogal,  Jutta and Ross,  Marta J. and Rumson,  Adrian F. and Sadiq,  Ghazala and Saeed,  Zeinab M. and Salimi,  Alireza and Salvalaglio,  Matteo and Sanders de Almada,  Leticia and Sasikumar,  Kiran and Sekharan,  Sivakumar and Shang,  Cheng and Shankland,  Kenneth and Shinohara,  Kotaro and Shi,  Baimei and Shi,  Xuekun and Skillman,  A. Geoffrey and Song,  Hongxing and Strasser,  Nina and van de Streek,  Jacco and Sugden,  Isaac J. and Sun,  Guangxu and Szalewicz,  Krzysztof and Tan,  Benjamin I. and Tan,  Lu and Tarczynski,  Frank and Taylor,  Christopher R. and Tkatchenko,  Alexandre and Tom,  Rithwik and Tuckerman,  Mark E. and Utsumi,  Yohei and Vogt-Maranto,  Leslie and Weatherston,  Jake and Wilkinson,  Luke J. and Willacy,  Robert D. and Wojtas,  Lukasz and Woollam,  Grahame R. and Yang,  Zhuocen and Yonemochi,  Etsuo and Yue,  Xin and Zeng,  Qun and Zhang,  Yizu and Zhou,  Tian and Zhou,  Yunfei and Zubatyuk,  Roman and Cole,  Jason C.},
  year = {2024},
  month = dec,
  pages = {517–547}
}

@article{Zhou2025,
  title = {A robust crystal structure prediction method to support small molecule drug development with large scale validation and blind study},
  volume = {16},
  ISSN = {2041-1723},
  url = {http://dx.doi.org/10.1038/s41467-025-57479-1},
  DOI = {10.1038/s41467-025-57479-1},
  number = {1},
  journal = {Nature Communications},
  publisher = {Springer Science and Business Media LLC},
  author = {Zhou,  Dong and Bier,  Imanuel and Santra,  Biswajit and Jacobson,  Leif D. and Wu,  Chuanjie and Garaizar Suarez,  Adiran and Almaguer,  Barbara Ramirez and Yu,  Haoyu and Abel,  Robert and Friesner,  Richard A. and Wang,  Lingle},
  year = {2025},
  month = mar 
}

@article{Kadan2023,
  title = {Accelerated Organic Crystal Structure Prediction with Genetic Algorithms and Machine Learning},
  volume = {19},
  ISSN = {1549-9626},
  url = {http://dx.doi.org/10.1021/acs.jctc.3c00853},
  DOI = {10.1021/acs.jctc.3c00853},
  number = {24},
  journal = {Journal of Chemical Theory and Computation},
  publisher = {American Chemical Society (ACS)},
  author = {Kadan,  Amit and Ryczko,  Kevin and Wildman,  Andrew and Wang,  Rodrigo and Roitberg,  Adrian and Yamazaki,  Takeshi},
  year = {2023},
  month = dec,
  pages = {9388–9402}
}

@article{greenwell2020,
author = {Greenwell, Chandler and Beran, Gregory J. O.},
title = {Inaccurate Conformational Energies Still Hinder Crystal Structure Prediction in Flexible Organic Molecules},
journal = {Crystal Growth \& Design},
volume = {20},
number = {8},
pages = {4875-4881},
year = {2020},
doi = {10.1021/acs.cgd.0c00676},
URL = { 
        https://doi.org/10.1021/acs.cgd.0c00676
},
eprint = { https://doi.org/10.1021/acs.cgd.0c00676}
}

@article{Beran2023,
  title = {Frontiers of molecular crystal structure prediction for pharmaceuticals and functional organic materials},
  volume = {14},
  ISSN = {2041-6539},
  url = {http://dx.doi.org/10.1039/D3SC03903J},
  DOI = {10.1039/d3sc03903j},
  number = {46},
  journal = {Chemical Science},
  publisher = {Royal Society of Chemistry (RSC)},
  author = {Beran,  Gregory J. O.},
  year = {2023},
  pages = {13290–13312}
}

@article{fenofibrate_IIa,
  title   = {Evidence of a metastable form of fenofibrate},
  author  = {Di Martino, P. and Palmieri, G. F. and Martelli, S.},
  journal = {Pharmazie},
  volume  = {55},
  number  = {8},
  pages   = {625--626},
  year    = {2000},
  issn    = {0031-7144},
  url     = {https://pubmed.ncbi.nlm.nih.gov/10989846/}
}

@article{Balendiran2012_fenofibrate,
  title   = {Biomolecular chemistry of isopropyl fibrates},
  author  = {Balendiran, Ganesaratnam K. and Rath, Niharika and Kotheimer, Amanda and Miller, Chad and Zeller, Matthias and Rath, Nigam P.},
  journal = {Journal of Pharmaceutical Sciences},
  volume  = {101},
  number  = {4},
  pages   = {1555--1569},
  year    = {2012},
  doi     = {10.1002/jps.23040},
  pmid    = {22246648},
  url     = {https://pubmed.ncbi.nlm.nih.gov/22246648/}
}

@patent{2018_fenofibrate_patent,
  title  = {Fenofibrate crystalline form and manufacturing method thereof},
  number = {WO2018018618A1},
  year   = {2018},
  author = {Shi, Xiangjun and Shao, Yinghua and Sheng, Xiaohong and Sheng, Xiaoxia},
  holder = {Zhejiang University of Technology; Hangzhou Lingye Pharmaceutical Technology Co.},
  note   = {Published 1 Feb 2018},
  url    = {https://patents.google.com/patent/WO2018018618A1/en}
}

@article{original_progesterone,
  title   = {Structure cristalline et moléculaire de la progestérone, C\_21H\_30O\_2},
  author  = {Campsteyn, H. and Dupont, L. and Dideberg, O.},
  journal = {Acta Crystallographica Section B},
  volume  = {28},
  pages   = {3032--3042},
  year    = {1972},
  doi     = {10.1107/S0567740872007393},
  url     = {https://doi.org/10.1107/S0567740872007393}
}

@article{racemic_progesterone,
  title   = {Racemic progesterone: predicted in silico and produced in the solid state},
  author  = {Lancaster, Robert W. and Karamertzanis, Panagiotis G. and Hulme, Ashley T. and Tocher, Derek A. and Covey, Douglas F. and Price, Sarah L.},
  journal = {Chemical Communications},
  year    = {2006},
  pages   = {4921--4923},
  doi     = {10.1039/B611599C},
  url     = {https://pubs.rsc.org/en/content/articlelanding/2006/cc/b611599c}
}

@article{Lee2014PolymorphGuide,
  title   = {A Practical Guide to Pharmaceutical Polymorph Screening & Selection},
  author  = {Lee, Eun Hee},
  journal = {Asian Journal of Pharmaceutical Sciences},
  volume  = {9},
  number  = {4},
  pages   = {163--175},
  year    = {2014},
  doi     = {10.1016/j.ajps.2014.05.002},
  url     = {https://doi.org/10.1016/j.ajps.2014.05.002}
}

@article{progesterone_formII,
  title   = {The polymorphism of progesterone: stabilization of a ‘disappearing’ polymorph by co-crystallization},
  author  = {Lancaster, Robert W. and Karamertzanis, Panagiotis G. and Hulme, Ashley T. and Tocher, Derek A. and Lewis, Thomas C. and Price, Sarah L.},
  journal = {Journal of Pharmaceutical Sciences},
  volume  = {96},
  number  = {12},
  pages   = {3419--3431},
  year    = {2007},
  doi     = {10.1002/jps.20983},
  url     = {https://pubmed.ncbi.nlm.nih.gov/17621678/}
}

@article{Nyman2015,
  title = {Static and lattice vibrational energy differences between polymorphs},
  volume = {17},
  ISSN = {1466-8033},
  url = {http://dx.doi.org/10.1039/C5CE00045A},
  DOI = {10.1039/c5ce00045a},
  number = {28},
  journal = {CrystEngComm},
  publisher = {Royal Society of Chemistry (RSC)},
  author = {Nyman,  Jonas and Day,  Graeme M.},
  year = {2015},
  pages = {5154–5165}
}

@article{Perdew2001,
    author = {Perdew, John P. and Schmidt, Karla},
    title = {Jacob’s ladder of density functional approximations for the exchange-correlation energy},
    journal = {AIP Conference Proceedings},
    volume = {577},
    number = {1},
    pages = {1-20},
    year = {2001},
    month = {07},
    issn = {0094-243X},
    doi = {10.1063/1.1390175},
    url = {https://doi.org/10.1063/1.1390175},
    eprint = {https://pubs.aip.org/aip/acp/article-pdf/577/1/1/12108089/1\_1\_online.pdf},
}

@article{PBE,
  title = {Generalized Gradient Approximation Made Simple},
  author = {Perdew, John P. and Burke, Kieron and Ernzerhof, Matthias},
  journal = {Phys. Rev. Lett.},
  volume = {77},
  issue = {18},
  pages = {3865--3868},
  numpages = {0},
  year = {1996},
  month = {Oct},
  publisher = {American Physical Society},
  doi = {10.1103/PhysRevLett.77.3865},
  url = {https://link.aps.org/doi/10.1103/PhysRevLett.77.3865}
}

@article{PBE0,
  title = {Toward reliable density functional methods without adjustable parameters: The PBE0 model},
  volume = {110},
  ISSN = {1089-7690},
  url = {http://dx.doi.org/10.1063/1.478522},
  DOI = {10.1063/1.478522},
  number = {13},
  journal = {The Journal of Chemical Physics},
  publisher = {AIP Publishing},
  author = {Adamo,  Carlo and Barone,  Vincenzo},
  year = {1999},
  month = apr,
  pages = {6158–6170}
}

@article{
Hoja2019,
author = {Johannes Hoja  and Hsin-Yu Ko  and Marcus A. Neumann  and Roberto Car  and Robert A. DiStasio  and Alexandre Tkatchenko },
title = {Reliable and practical computational description of molecular crystal polymorphs},
journal = {Science Advances},
volume = {5},
number = {1},
pages = {eaau3338},
year = {2019},
doi = {10.1126/sciadv.aau3338},
URL = {https://www.science.org/doi/abs/10.1126/sciadv.aau3338},
eprint = {https://www.science.org/doi/pdf/10.1126/sciadv.aau3338},
}

@article{Agarwal2022,
title = {Trends in small molecule drug properties: A developability molecule assessment perspective},
journal = {Drug Discovery Today},
volume = {27},
number = {12},
pages = {103366},
year = {2022},
issn = {1359-6446},
doi = {https://doi.org/10.1016/j.drudis.2022.103366},
url = {https://www.sciencedirect.com/science/article/pii/S1359644622003592},
author = {Prashant Agarwal and James Huckle and Jake Newman and Darren L. Reid},
}

@book{martin2020,
  title={Electronic Structure: Basic Theory and Practical Methods},
  author={Martin, R.M.},
  isbn={9781108657471},
  year={2020},
  publisher={Cambridge University Press}
}

@article{GFN2,
author = {Bannwarth, Christoph and Ehlert, Sebastian and Grimme, Stefan},
title = {GFN2-xTB—An Accurate and Broadly Parametrized Self-Consistent Tight-Binding Quantum Chemical Method with Multipole Electrostatics and Density-Dependent Dispersion Contributions},
journal = {Journal of Chemical Theory and Computation},
volume = {15},
number = {3},
pages = {1652-1671},
year = {2019},
doi = {10.1021/acs.jctc.8b01176},
note ={PMID: 30741547},
URL = { https://doi.org/10.1021/acs.jctc.8b01176},
eprint = { https://doi.org/10.1021/acs.jctc.8b01176}
}

@misc{IUCR2016,
  title = {International Tables for Crystallography: Space-group symmetry},
  ISBN = {9780470974230},
  url = {http://dx.doi.org/10.1107/97809553602060000114},
  DOI = {10.1107/97809553602060000114},
  journal = {International Tables for Crystallography},
  publisher = {International Union of Crystallography},
  year = {2016},
  month = dec 
}

@article{Duignan2024,
  title = {The Potential of Neural Network Potentials},
  volume = {4},
  ISSN = {2694-2445},
  url = {http://dx.doi.org/10.1021/acsphyschemau.4c00004},
  DOI = {10.1021/acsphyschemau.4c00004},
  number = {3},
  journal = {ACS Physical Chemistry Au},
  publisher = {American Chemical Society (ACS)},
  author = {Duignan,  Timothy T.},
  year = {2024},
  month = mar,
  pages = {232–241}
}

@article{Kocer2022,
  title = {Neural Network Potentials: A Concise Overview of Methods},
  volume = {73},
  ISSN = {1545-1593},
  url = {http://dx.doi.org/10.1146/annurev-physchem-082720-034254},
  DOI = {10.1146/annurev-physchem-082720-034254},
  number = {1},
  journal = {Annual Review of Physical Chemistry},
  publisher = {Annual Reviews},
  author = {Kocer,  Emir and Ko,  Tsz Wai and Behler,  J\"{o}rg},
  year = {2022},
  month = apr,
  pages = {163–186}
}

@article{Gomez2024,
  title = {Neural-network-based molecular dynamics simulations reveal that proton transport in water is doubly gated by sequential hydrogen-bond exchange},
  volume = {16},
  ISSN = {1755-4349},
  url = {http://dx.doi.org/10.1038/s41557-024-01593-y},
  DOI = {10.1038/s41557-024-01593-y},
  number = {11},
  journal = {Nature Chemistry},
  publisher = {Springer Science and Business Media LLC},
  author = {Gomez,  Axel and Thompson,  Ward H. and Laage,  Damien},
  year = {2024},
  month = aug,
  pages = {1838–1844}
}

@InProceedings{Gilmer2017,
  title = 	 {Neural Message Passing for Quantum Chemistry},
  author =       {Justin Gilmer and Samuel S. Schoenholz and Patrick F. Riley and Oriol Vinyals and George E. Dahl},
  booktitle = 	 {Proceedings of the 34th International Conference on Machine Learning},
  pages = 	 {1263--1272},
  year = 	 {2017},
  editor = 	 {Precup, Doina and Teh, Yee Whye},
  volume = 	 {70},
  series = 	 {Proceedings of Machine Learning Research},
  month = 	 {06--11 Aug},
  publisher =    {PMLR},
  url = 	 {https://proceedings.mlr.press/v70/gilmer17a.html}
}

@article{AP-Net2,
  title = {A physics-aware neural network for protein–ligand interactions with quantum chemical accuracy},
  volume = {15},
  ISSN = {2041-6539},
  url = {http://dx.doi.org/10.1039/D4SC01029A},
  DOI = {10.1039/d4sc01029a},
  number = {33},
  journal = {Chemical Science},
  publisher = {Royal Society of Chemistry (RSC)},
  author = {Glick,  Zachary L. and Metcalf,  Derek P. and Glick,  Caroline S. and Spronk,  Steven A. and Koutsoukas,  Alexios and Cheney,  Daniel L. and Sherrill,  C. David},
  year = {2024},
  pages = {13313–13324}
}

@article{AP-Net,
  title = {AP-Net: An atomic-pairwise neural network for smooth and transferable interaction potentials},
  volume = {153},
  ISSN = {1089-7690},
  url = {http://dx.doi.org/10.1063/5.0011521},
  DOI = {10.1063/5.0011521},
  number = {4},
  journal = {The Journal of Chemical Physics},
  publisher = {AIP Publishing},
  author = {Glick,  Zachary L. and Metcalf,  Derek P. and Koutsoukas,  Alexios and Spronk,  Steven A. and Cheney,  Daniel L. and Sherrill,  C. David},
  year = {2020},
  month = jul 
}

@article{verstraelen2016minimal,
  title={Minimal basis iterative stockholder: atoms in molecules for force-field development},
  author={Verstraelen, Toon and Vandenbrande, Steven and Heidar-Zadeh, Farnaz and Vanduyfhuys, Louis and Van Speybroeck, Veronique and Waroquier, Michel and Ayers, Paul W},
  journal={Journal of Chemical Theory and Computation},
  volume={12},
  number={8},
  pages={3894--3912},
  year={2016},
  publisher={ACS Publications}
}

@article{Neumann2008,
  title = {Tailor-Made Force Fields for Crystal-Structure Prediction},
  volume = {112},
  ISSN = {1520-5207},
  url = {http://dx.doi.org/10.1021/jp710575h},
  DOI = {10.1021/jp710575h},
  number = {32},
  journal = {The Journal of Physical Chemistry B},
  publisher = {American Chemical Society (ACS)},
  author = {Neumann,  Marcus A.},
  year = {2008},
  month = jul,
  pages = {9810–9829}
}

@article{Bryenton2022,
  title = {Delocalization error: The greatest outstanding challenge in density‐functional theory},
  volume = {13},
  ISSN = {1759-0884},
  url = {http://dx.doi.org/10.1002/wcms.1631},
  DOI = {10.1002/wcms.1631},
  number = {2},
  journal = {WIREs Computational Molecular Science},
  publisher = {Wiley},
  author = {Bryenton,  Kyle R. and Adeleke,  Adebayo A. and Dale,  Stephen G. and Johnson,  Erin R.},
  year = {2022},
  month = jul 
}

@article{Urzhumtseva2009,
author = "Urzhumtseva, Ludmila and Afonine, Pavel V. and Adams, Paul D. and Urzhumtsev, Alexandre",
title = "{Crystallographic model quality at a glance}",
journal = "Acta Crystallographica Section D",
year = "2009",
volume = "65",
number = "3",
pages = "297--300",
month = "Mar",
doi = {10.1107/S0907444908044296},
url = {https://doi.org/10.1107/S0907444908044296},
}

@article{Newman2022Powders,
  author  = {Newman, Justin A. and Iuzzolino, Luca and Tan, Melissa and
             Orth, Peter and Bruhn, Jessica and Lee, Alfred Y.},
  title   = {From Powders to Single Crystals: A Crystallographer’s Toolbox
             for Small-Molecule Structure Determination},
  journal = {Molecular Pharmaceutics},
  year    = {2022},
  volume  = {19},
  number  = {7},
  pages   = {2133--2141},
  doi     = {10.1021/acs.molpharmaceut.2c00020}
}

@article{OterodelaRoza2024PXRD,
  author  = {Otero-de-la-Roza, Alberto},
  title   = {Powder-Diffraction-Based Structural Comparison for Crystal Structure Prediction without Prior Indexing},
  journal = {ChemRxiv},
  year    = {2024},
  doi     = {10.26434/chemrxiv-2024-hdt4m-v2}
}

@patent{omaveloxolone_forms,
  author   = {Ahmad Y. Sheikh and Alessandra Mattei and Xiu C. Wang},
  title    = {2,2-Difluoropropionamide Derivatives of Bardoxolone Methyl, Polymorphic Forms and Methods of Use Thereof},
  number   = {US 2020/0062800 A1},
  year     = {2020},
  month    = {February},
  day      = {27},
  assignee = {AbbVie Inc.},
  url      = {https://patents.google.com/patent/US20200062800A1/en}
}

@article{omaveloxolone_general,
  author = {Pilotto, Fiammetta and Chellapandi, Deepika and Puccio, Hélène},
  title = {Omaveloxolone: a groundbreaking milestone as the first FDA-approved drug for Friedreich ataxia},
  journal = {Trends in Molecular Medicine},
  volume = {30},
  number = {2},
  pages = {117--125},
  year = {2024},
  month = {feb},
  doi = {10.1016/j.molmed.2023.12.002},
  url = {https://doi.org/10.1016/j.molmed.2023.12.002},
  pmid = {38272714}
}

@patent{zuranolone_forms,
  author    = {Watson, Paul S. and Berner, Bret and Reid, John G. and Wang, Jian and Doherty, James and Kanes, Stephen J.},
  title     = {A crystalline 19-nor C\textsubscript{3,3}-disubstituted C\textsubscript{21}-N-pyrazolyl steroid},
  number    = {WO2018039378A1},
  assignee  = {Sage Therapeutics, Inc.},
  year      = {2018},
  month     = mar,
  day       = {1},
  url       = {https://patentimages.storage.googleapis.com/55/f0/c9/fe279911c9159f/WO2018039378A1.pdf}
}

@article{zuranolone_general,
  author  = {Heo, Young-A},
  title   = {Zuranolone: First Approval},
  journal = {Drugs},
  year    = {2023},
  volume  = {83},
  number  = {16},
  pages   = {1559--1567},
  month   = nov,
  doi     = {10.1007/s40265-023-01953-x},
  pmid    = {37882942}
}

@article{deucravacitinib_general,
  author  = {Hoy, Sheridan M.},
  title   = {Deucravacitinib: First Approval},
  journal = {Drugs},
  year    = {2022},
  volume  = {82},
  number  = {17},
  pages   = {1671--1679},
  month   = nov,
  doi     = {10.1007/s40265-022-01796-y},
  pmid    = {36401743},
  pmcid   = {PMC9676857}
}

@patent{deucravacitinib_1,
  author    = {Yanlei Zhang and Michael A. Galella},
  title     = {Crystal form of 6-(cyclopropanecarboxamido)-4-((2-methoxy-3-(1-methyl-1H-1,2,4-triazol-3-yl)phenyl)amino)-N-(methyl-d3) pyridazine-3-carboxamide},
  number    = {WO 2018/183656 A1},
  assignee  = {Bristol-Myers Squibb Company},
  year      = {2018},
  month     = {Oct},
  day       = {4},
}

@patent{deucravacitinib_2,
  author    = {Daniel Richard Roberts},
  title     = {Crystalline form of 6-(cyclopropanecarboxamido)-4-((2-methoxy-3-(1-methyl-1H-1,2,4-triazol-3-yl)phenyl)amino)-N-(methyl-d3) pyridazine-3-carboxamide},
  number    = {WO 2019/232138 A1},
  assignee  = {Bristol-Myers Squibb Company},
  year      = {2019},
  month     = {Dec},
  day       = {5},
}

@patent{deucravacitinib_3,
  author    = {Daniel Richard Roberts and Chenkou Wei},
  title     = {Crystalline salt forms of 6-(cyclopropanecarboxamido)-4-((2-methoxy-3-(1-methyl-1H-1,2,4-triazol-3-yl)phenyl)amino)-N-(methyl-d3) pyridazine-3-carboxamide},
  number    = {WO 2020/251911 A1},
  assignee  = {Bristol-Myers Squibb Company},
  year      = {2020},
  month     = {Dec},
  day       = {17},
}

@patent{deucravacitinib_4,
  author    = {Minhua Chen and Hongyan Zhu},
  title     = {BMS-986165 crystal form, preparation method therefor and use thereof},
  number    = {WO 2021/129467 A1},
  assignee  = {Suzhou Keruisi Pharmaceutical Co., Ltd.},
  year      = {2021},
  month     = {Jul},
  day       = {1},
}

@patent{deucravacitinib_5,
  author    = {Minhua Chen and Hongyan Zhu},
  title     = {Deucravacitinib crystal form, preparation method therefor and use thereof},
  number    = {WO 2021/143498 A1},
  assignee  = {Suzhou Keruisi Pharmaceutical Co., Ltd.},
  year      = {2021},
  month     = {Jul},
  day       = {22},
}

@patent{deucravacitinib_6,
  author    = {Victor W. Rosso},
  title     = {Crystal form of 6-(cyclopropanecarboxamido)-4-((2-methoxy-3-(1-methyl-1H-1,2,4-triazol-3-yl)phenyl)amino)-N-(methyl-d3) pyridazine-3-carboxamide},
  number    = {US 12,129,245 B2},
  assignee  = {Bristol Myers Squibb Company},
  year      = {2024},
  month     = {Oct},
  day       = {29},
}

@patent{deucravacitinib_7,
  author    = {Candice Y. Choi and Daniel Richard Roberts and Chenkou Wei and Marta Dabros and Franz Lembke and Ian Yates and Natalie Louise Kelk and David James Pearson},
  title     = {Crystal forms of 6-(cyclopropanecarboxamido)-4-((2-methoxy-3-(1-methyl-1H-1,2,4-triazol-3-yl)phenyl)amino)-N-(methyl-d3)pyridazine-3-carboxamide},
  number    = {US 2024/0190845 A1},
  assignee  = {Bristol Myers Squibb Company},
  year      = {2024},
  month     = {Jun},
  day       = {13},
}

@misc{deucravacitinib_8,
  author    = {Srinivasan Thirumalai Rajan and Vijayavitthal T. Mathad and Ismail and Nevuluri Narasimha Rao},
  title     = {Process for the preparation of amorphous form of BMS-986165},
  howpublished = {Technical Disclosure Commons Defensive Publication},
  institution = {MSN Laboratories Private Limited, R\&D Center},
  year      = {2024},
  month     = {Oct},
  day       = {25},
}

@patent{deucravacitinib_9,
  author    = {Arijit Das and Ramanaiah Chennuru and Ramesh Devarapalli and Anjaneyaraju Indukuri and Manjunath Bollineni},
  title     = {Deucravacitinib amorphous solid dispersions and polymorphs thereof},
  number    = {WO 2024/176263 A1},
  assignee  = {Cipla Limited},
  year      = {2024},
  month     = {Aug},
  day       = {29}
}

@article{Podryabinkin2019,
  author  = {Podryabinkin, Evgeny V. and Tikhonov, Evgeny V. and Shapeev, Alexander V. and Oganov, Artem R.},
  title   = {Accelerating Crystal Structure Prediction by Machine-Learning Interatomic Potentials with Active Learning},
  journal = {Physical Review B},
  year    = {2019},
  volume  = {99},
  pages   = {064114},
  doi     = {10.1103/PhysRevB.99.064114}
}

@article{Butler2024,
  author  = {Butler, Patrick W. V. and Hafizi, Roohollah and Day, Graeme M.},
  title   = {Machine-Learned Potentials by Active Learning from Organic Crystal Structure Prediction Landscapes},
  journal = {Journal of Physical Chemistry A},
  year    = {2024},
  volume  = {128},
  number  = {5},
  pages   = {945--957},
  doi     = {10.1021/acs.jpca.3c07129}
}

@article{Taylor2024,
  author  = {Taylor, Christopher R. and Butler, Patrick W. V. and Day, Graeme M.},
  title   = {Predictive Crystallography at Scale: Mapping, Validating, and Learning from 1000 Crystal Energy Landscapes},
  journal = {Faraday Discussions},
  year    = {2025},
  volume  = {256},
  pages   = {434--458},
  doi     = {10.1039/d4fd00105b}
}

@article{Omee2024,
  author  = {Omee, Sadman Sadeed and Wei, Lai and Hu, Jianjun},
  title   = {Crystal Structure Prediction Using Neural Network Potential and Age-Fitness Pareto Genetic Algorithm},
  journal = {Journal of Materials Informatics},
  year    = {2024},
  volume  = {4},
  pages   = {2},
  doi     = {10.20517/jmi.2023.33}
}

@article{Rybin2025,
  author  = {Rybin, Nikita and Novikov, Ivan S. and Shapeev, Alexander},
  title   = {Accelerating Structure Prediction of Molecular Crystals Using Actively Trained Moment Tensor Potential},
  journal = {Physical Chemistry Chemical Physics},
  year    = {2025},
  volume  = {27},
  pages   = {5141--5148},
  doi     = {10.1039/D4CP04578E}
}

@article{chlorpropamide,
  title = {Multicomponent Crystals of Chlorpropamide: Multiple Conformers,  Multiple Z′,  and Proton Transfer at Play},
  volume = {21},
  ISSN = {1528-7505},
  url = {http://dx.doi.org/10.1021/acs.cgd.1c00236},
  DOI = {10.1021/acs.cgd.1c00236},
  number = {6},
  journal = {Crystal Growth \& Design},
  publisher = {American Chemical Society (ACS)},
  author = {B,  Haripriya and Hasija,  Avantika and Cruz-Cabeza,  Aurora J. and Shruti,  Ipsha and Chopra,  Deepak},
  year = {2021},
  month = may,
  pages = {3158–3167}
}

@article{tolbutamide,
  title = {Conformational Polymorphism of Tolbutamide: A Structural,  Spectroscopic,  and Thermodynamic Characterization of Burger’s Forms I–IV},
  volume = {99},
  ISSN = {0022-3549},
  url = {http://dx.doi.org/10.1002/jps.22061},
  DOI = {10.1002/jps.22061},
  number = {7},
  journal = {Journal of Pharmaceutical Sciences},
  publisher = {Elsevier BV},
  author = {Thirunahari,  Satyanarayana and Aitipamula,  Srinivasulu and Chow,  Pui Shan and Tan,  Reginald B.H.},
  year = {2010},
  month = jul,
  pages = {2975–2990}
}

@article{sildenafil,
author = {Barbas, Rafael and Font-Bardia, Mercè and Prohens, Rafel},
title = {Polymorphism of Sildenafil: A New Metastable Desolvate},
journal = {Crystal Growth \& Design},
volume = {18},
number = {7},
pages = {3740-3746},
year = {2018},
doi = {10.1021/acs.cgd.8b00683},
URL = {https://doi.org/10.1021/acs.cgd.8b00683},
eprint = {https://doi.org/10.1021/acs.cgd.8b00683}
}

@patent{rotigotine_form_1,
  author = {Wolff. Hans-Michael and Quere, Luc and Riedner, Jens},
  title = {Polymorphic form of rotigotine and process for production},
  number = {US8232414B2},
  year = {2012},
  month = {July},
  day = {31},
  organization = {United States Patent and Trademark Office},
  note = {Issued},
}

@article{rotigotine,
title = {Rotigotine: The first new chemical entity for transdermal drug delivery},
journal = {European Journal of Pharmaceutics and Biopharmaceutics},
volume = {88},
number = {3},
pages = {586-593},
year = {2014},
issn = {0939-6411},
doi = {https://doi.org/10.1016/j.ejpb.2014.08.007},
url = {https://www.sciencedirect.com/science/article/pii/S0939641114002483},
author = {Donald A. McAfee and Jonathan Hadgraft and Majella E. Lane}
}

@article{Chaudhuri01112008,
author = {K Ray Chaudhuri},
title = {Crystallisation within transdermal rotigotine patch: is there cause for concern?},
journal = {Expert Opinion on Drug Delivery},
volume = {5},
number = {11},
pages = {1169--1171},
year = {2008},
publisher = {Taylor \& Francis},
doi = {10.1517/17425240802500870},
note ={PMID: 18976128},
URL = {https://doi.org/10.1517/17425240802500870},
eprint = {https://doi.org/10.1517/17425240802500870}
}

@article{Mortazavi2019,
  title = {Computational polymorph screening reveals late-appearing and poorly-soluble form of rotigotine},
  volume = {2},
  ISSN = {2399-3669},
  url = {http://dx.doi.org/10.1038/s42004-019-0171-y},
  DOI = {10.1038/s42004-019-0171-y},
  number = {1},
  journal = {Communications Chemistry},
  publisher = {Springer Science and Business Media LLC},
  author = {Mortazavi,  Majid and Hoja,  Johannes and Aerts,  Luc and Quéré,  Luc and van de Streek,  Jacco and Neumann,  Marcus A. and Tkatchenko,  Alexandre},
  year = {2019},
  month = jun 
}

@article{Rietveld2015,
  title = {Rotigotine: Unexpected Polymorphism with Predictable Overall Monotropic Behavior},
  volume = {104},
  ISSN = {0022-3549},
  url = {http://dx.doi.org/10.1002/jps.24626},
  DOI = {10.1002/jps.24626},
  number = {12},
  journal = {Journal of Pharmaceutical Sciences},
  publisher = {Elsevier BV},
  author = {Rietveld,  Ivo B. and Céolin,  René},
  year = {2015},
  month = dec,
  pages = {4117–4122}
}

@article{Hughes2011,
  author    = {Hughes, J.P. and Rees, S. and Kalindjian, S.B. and Philpott, K.L.},
  title     = {Principles of early drug discovery},
  journal   = {British Journal of Pharmacology},
  year      = {2011},
  volume    = {162},
  number    = {6},
  pages     = {1239--1249},
  doi       = {10.1111/j.1476-5381.2010.01127.x},
  url       = {https://bpspubs.onlinelibrary.wiley.com/doi/full/10.1111/j.1476-5381.2010.01127.x}
}

@article{Yang2022MetadynamicsNNP,
  author  = {Yang, Manyi and Bonati, Luigi and Polino, Daniela and Parrinello, Michele},
  title   = {Using metadynamics to build neural network potentials for reactive events: the case of urea decomposition in water},
  journal = {Catalysis Today},
  year    = {2022},
  volume  = {387},
  pages   = {143--149},
  doi     = {10.1016/j.cattod.2021.03.018}
}

@article{Anstine2025AIMNet2,
  author  = {Anstine, Dylan M. and Zubatyuk, Roman and Isayev, Olexandr},
  title   = {AIMNet2: a Neural Network Potential to Meet Your Neutral, Charged, Organic, and Elemental-Organic Needs},
  journal = {Chemical Science},
  year    = {2025},
  volume  = {16},
  pages   = {10228--10244},
  month   = apr,
  doi     = {10.1039/D4SC08572H}
}

@article{SabanesZariquiey2024NNPBinding,
  author  = {Sabanés Zariquiey, Francesc and Galvelis, Raimondas and Gallicchio, Emilio
             and Chodera, John D. and Markland, Thomas E. and De Fabritiis, Gianni},
  title   = {Enhancing Protein–Ligand Binding Affinity Predictions Using Neural Network Potentials},
  journal = {Journal of Chemical Information and Modeling},
  year    = {2024},
  volume  = {64},
  number  = {5},
  pages   = {1481--1485},
  doi     = {10.1021/acs.jcim.3c02031}
}

@article{Lian2024ODCu,
  author  = {Lian, Zan and Dattila, Federico and López, Núria},
  title   = {Stability and Lifetime of Diffusion-Trapped Oxygen in Oxide-Derived Copper CO\textsubscript{2} Reduction Electrocatalysts},
  journal = {Nature Catalysis},
  year    = {2024},
  volume  = {7},
  pages   = {401--411},
  doi     = {10.1038/s41929-024-01132-5}
}

@article{Cao2025ZincElectrolyte,
  author  = {Cao, Chuntian and Kingan, Arun and Hill, Ryan C. and Kuang, Jason and Wang, Lei
             and Zhang, Chunyi and Carbone, Matthew R. and van Dam, Hubertus and Yoo, Shinjae
             and Marschilok, Amy C. and Lu, Deyu},
  title   = {Resolving the Solvation Structure and Transport Properties of Aqueous Zinc Electrolytes from Salt-in-Water to Water-in-Salt Using Neural Network Potential},
  journal = {PRX Energy},
  year    = {2025},
  volume  = {4},
  pages   = {023004},
  doi     = {10.1103/PRXEnergy.4.023004}
}

@article{Takamoto2022PFP,
  author  = {Takamoto, So and Shinagawa, Chikashi and Motoki, Daisuke and Nakago, Kosuke
             and Li, Wenwen and Kurata, Iori and Watanabe, Taku \emph{et al.}},
  title   = {Towards Universal Neural Network Potential for Material Discovery Applicable to Arbitrary Combination of 45 Elements},
  journal = {Nature Communications},
  year    = {2022},
  volume  = {13},
  pages   = {2991},
  doi     = {10.1038/s41467-022-30687-9}
}

@article{Ko2021Accounts,
  author  = {Ko, Tsz Wai and Finkler, Jonas A. and Goedecker, Stefan and Behler, Jörg},
  title   = {General-Purpose Machine Learning Potentials Capturing Nonlocal Charge Transfer},
  journal = {Accounts of Chemical Research},
  year    = {2021},
  volume  = {54},
  number  = {4},
  pages   = {808--817},
  doi     = {10.1021/acs.accounts.0c00689}
}

@article{Musaelian2023Allegro,
  author  = {Musaelian, Albert and Batzner, Simon and Johansson, Anders and Sun, Lixin and Owen, Cameron J. and Kornbluth, Mordechai and Kozinsky, Boris},
  title   = {Learning Local Equivariant Representations for Large-Scale Atomistic Dynamics},
  journal = {Nature Communications},
  year    = {2023},
  volume  = {14},
  pages   = {579},
  month   = feb,
  doi     = {10.1038/s41467-023-36329-y}
}

@article{Borca2023MBE,
  author  = {Borca, Carlos H. and Glick, Zachary L. and Metcalf, Derek P. and Burns, Lori A. and Sherrill, C. David},
  title   = {Benchmark Coupled‐Cluster Lattice Energy of Crystalline Benzene and Assessment of Multi‐Level Approximations in the Many‐Body Expansion},
  journal = {Journal of Chemical Physics},
  year    = {2023},
  volume  = {158},
  pages   = {234102},
  month   = jun,
  doi     = {10.1063/5.0159410}
}

@article{Kennedy2014ThreeBody,
  author  = {Kennedy, Matthew R. and McDonald, Ashley Ringer and DePrince, A. Eugene III and Marshall, Michael S. and Podeszwa, Rafal and Sherrill, C. David},
  title   = {Communication: Resolving the Three‐Body Contribution to the Lattice Energy of Crystalline Benzene: Benchmark Results from Coupled‐Cluster Theory},
  journal = {Journal of Chemical Physics},
  year    = {2014},
  volume  = {140},
  pages   = {121104},
  month   = mar,
  doi     = {10.1063/1.4869686}
}

@article{Wen2011FragmentQM,
  author  = {Wen, Shuhao and Beran, Gregory J. O.},
  title   = {Accurate Molecular Crystal Lattice Energies from a Fragment QM/MM Approach with On‐the‐Fly \emph{Ab Initio} Force Field Parametrization},
  journal = {Journal of Chemical Theory and Computation},
  year    = {2011},
  volume  = {7},
  number  = {11},
  pages   = {3733--3742},
  month   = nov,
  doi     = {10.1021/ct200541h}
}

@article{Herman2023IceMBE,
  author  = {Herman, Kristina M. and Xantheas, Sotiris S.},
  title   = {A Formulation of the Many‐Body Expansion (MBE) for Periodic Systems: Application to Several Ice Phases},
  journal = {Journal of Physical Chemistry Letters},
  year    = {2023},
  volume  = {14},
  number  = {4},
  pages   = {989--999},
  month   = feb,
  doi     = {10.1021/acs.jpclett.2c03822}
}

@article{Bannan2025CSPCloud,
  author  = {Bannan, Caitlin C. and Ovanesyan, Grigory and Darden, Thomas A. and Graves, Andrew P. and Edge, Christopher M. and Russo, Lucas and Davenport, Aidan P. \emph{et al.}},
  title   = {Crystal Structure Prediction of Drug Molecules in the Cloud: A Collaborative Blind Challenge Study},
  journal = {Crystal Growth \& Design},
  year    = {2025},
  volume  = {25},
  number  = {5},
  pages   = {1299--1314},
  month   = may,
  doi     = {10.1021/acs.cgd.4c00572}
}

@article{Smith2020Psi4,
  author  = {Smith, Daniel G. A. and Burns, Lori A. and Simmonett, Andrew C.
             and Parrish, Robert M. and Schieber, Matthew C. and Galvelis, Raimondas
             and Kraus, Peter and Kruse, Holger and Di Remigio, Roberto
             and others},
  title   = {Psi4 1.4: Open-Source Software for High-Throughput Quantum Chemistry},
  journal = {Journal of Chemical Physics},
  year    = {2020},
  volume  = {152},
  number  = {18},
  pages   = {184108},
  doi     = {10.1063/5.0006002}
}

@article{Mueller2023wB97X3c,
  author  = {Müller, Marcel and Hansen, Andreas and Grimme, Stefan},
  title   = {ωB97X-3c: A Composite Range‐Separated Hybrid DFT Method with a Molecule‐Optimized Polarized Valence Double‐ζ Basis Set},
  journal = {Journal of Chemical Physics},
  year    = {2023},
  volume  = {158},
  number  = {1},
  pages   = {014103},
  doi     = {10.1063/5.0133026}
}

@article{Chai2008wB97,
  author  = {Chai, Jeng‐Da and Head‐Gordon, Martin},
  title   = {Systematic Optimization of Long‐Range Corrected Hybrid Density Functionals},
  journal = {Journal of Chemical Physics},
  year    = {2008},
  volume  = {128},
  pages   = {084106},
  doi     = {10.1063/1.2834918}
}

@article{Grimme2010D3,
  author  = {Grimme, Stefan and Antony, Jan and Ehrlich, Stephan and Krieg, Helge},
  title   = {A Consistent and Accurate \emph{Ab Initio} Parametrization of Density‐Functional Dispersion Correction (DFT‐D) for the 94 Elements H–Pu},
  journal = {Journal of Chemical Physics},
  year    = {2010},
  volume  = {132},
  pages   = {154104},
  doi     = {10.1063/1.3382344}
}

@article{Grimme2011BJ,
  author  = {Grimme, Stefan and Ehrlich, Stephan and Goerigk, Lars},
  title   = {Effect of the Damping Function in Dispersion‐Corrected Density Functional Theory},
  journal = {Journal of Computational Chemistry},
  year    = {2011},
  volume  = {32},
  pages   = {1456--1465},
  doi     = {10.1002/jcc.21759}
}

@article{Weigend2005def2,
  author  = {Weigend, Florian and Ahlrichs, Reinhart},
  title   = {Balanced Basis Sets of Split Valence, Triple Zeta Valence and Quadruple Zeta Valence Quality for H to Rn: Design and Assessment of Accuracy},
  journal = {Physical Chemistry Chemical Physics},
  year    = {2005},
  volume  = {7},
  pages   = {3297--3305},
  doi     = {10.1039/B508541A}
}

@article{Hellweg2015TZVPPD,
  author  = {Hellweg, Arnim and Rappoport, Dmitrij},
  title   = {Development of New Auxiliary Basis Functions of the Karlsruhe Segmented Contracted Basis Sets Including Diffuse Basis Functions (def2‐SVPD, def2‐TZVPPD, and def2‐QZVPPD) for RI‐MP2 and RI‐CC Calculations},
  journal = {Physical Chemistry Chemical Physics},
  year    = {2015},
  volume  = {17},
  pages   = {1010--1017},
  doi     = {10.1039/C4CP04286G}
}

@article{Bravetti2022Mebendazole,
  author  = {Bravetti, Federica and Bordignon, Simone and Alig, Edith and
             Eisenbeil, Daniel and Fink, Lothar and Nervi, Carlo and Gobetto, Roberto and
             Schmidt, Martin U. and Chierotti, Michele R.},
  title   = {Solid-State NMR-Driven Crystal Structure Prediction of Molecular Crystals:
             The Case of Mebendazole},
  journal = {Chemistry – A European Journal},
  year    = {2022},
  volume  = {28},
  number  = {6},
  pages   = {e202103589},
  doi     = {10.1002/chem.202103589}
}

@book{Sidhu2023Fenofibrate,
  author    = {Sidhu, Gursharan and Tripp, Jayson},
  title     = {Fenofibrate},
  booktitle = {StatPearls [Internet]},
  publisher = {StatPearls Publishing},
  address   = {Treasure Island (FL)},
  year      = {2023},
  note      = {Last update March 13, 2023},
  url       = {https://www.ncbi.nlm.nih.gov/books/NBK559219/}
}

@article{Frampton2019Rotigotine,
  author  = {Frampton, James E.},
  title   = {Rotigotine Transdermal Patch: A Review in Parkinson's Disease},
  journal = {CNS Drugs},
  year    = {2019},
  volume  = {33},
  number  = {7},
  pages   = {707--718},
  doi     = {10.1007/s40263-019-00646-y},
  pmid    = {31243728}
}

@article{Zhang2019,
  title = {Active learning of uniformly accurate interatomic potentials for materials simulation},
  author = {Zhang, Linfeng and Lin, De-Ye and Wang, Han and Car, Roberto and E, Weinan},
  journal = {Phys. Rev. Mater.},
  volume = {3},
  issue = {2},
  pages = {023804},
  numpages = {9},
  year = {2019},
  month = {Feb},
  publisher = {American Physical Society},
  doi = {10.1103/PhysRevMaterials.3.023804},
  url = {https://link.aps.org/doi/10.1103/PhysRevMaterials.3.023804}
}

@article{Bhardwaj2019Galunisertib,
  author  = {Bhardwaj, Rajni M. and McMahon, Jennifer A. and Nyman, Jonas
             and Price, Louise S. and Konar, Sumit and Oswald, Iain D. H.
             and Pulham, Colin R. and Price, Sarah L. and Reutzel-Edens, Susan M.},
  title   = {A Prolific Solvate Former, Galunisertib, under the Pressure of
             Crystal Structure Prediction, Produces Ten Diverse Polymorphs},
  journal = {Journal of the American Chemical Society},
  year    = {2019},
  volume  = {141},
  number  = {35},
  pages   = {13887--13897},
  doi     = {10.1021/jacs.9b06634}
}

@article{orbital-ritonavir,
  author  = {Bauser, Haley C. and Smith, Pamela A. and Parent, Stephan D. and Chan, Larry R. and Bhavsar, Ami S. and Condon, Kenneth H. and McCalip, Andrew and Croom, Jordan M. and Purcell, Dale K. and Bogdanowich-Knipp, Susan J. and Smith, Daniel T. and Cowans, Brett A. and Alajlouni, Ruba and Byrn, Stephen R. and Radocea, Adrian},
  title   = {Return of the Ritonavir: A Study on the Stability of Pharmaceuticals Processed in Orbit and Returned to Earth},
  journal = {ChemRxiv},
  year    = {2024},
  doi     = {10.26434/chemrxiv-2024-vb20g-v3}
}

@article{mace-off,
author = {Kovács, Dávid P{\'e}ter and Moore, J. Harry and Browning, Nicholas J. and Batatia, Ilyes and Horton, Joshua T. and Pu, Yixuan and Kapil, Venkat and Witt, William C. and Magdău, Ioan-Bogdan and Cole, Daniel J. and Csányi, Gábor},
title = {MACE-OFF: Short-Range Transferable Machine Learning Force Fields for Organic Molecules},
journal = {Journal of the American Chemical Society},
volume = {147},
number = {21},
pages = {17598-17611},
year = {2025},
doi = {10.1021/jacs.4c07099},
note ={PMID: 40387214},
URL = {https://doi.org/10.1021/jacs.4c07099},
eprint = {https://doi.org/10.1021/jacs.4c07099}

}

@misc{egret1,
      title={Egret-1: Pretrained Neural Network Potentials for Efficient and Accurate Bioorganic Simulation}, 
      author={Elias L. Mann and Corin C. Wagen and Jonathon E. Vandezande and Arien M. Wagen and Spencer C. Schneider},
      year={2025},
      eprint={2504.20955},
      archivePrefix={arXiv},
      primaryClass={physics.chem-ph},
      url={https://arxiv.org/abs/2504.20955}, 
}

@misc{uma,
      title={UMA: A Family of Universal Models for Atoms}, 
      author={Brandon M. Wood and Misko Dzamba and Xiang Fu and Meng Gao and Muhammed Shuaibi and Luis Barroso-Luque and Kareem Abdelmaqsoud and Vahe Gharakhanyan and John R. Kitchin and Daniel S. Levine and Kyle Michel and Anuroop Sriram and Taco Cohen and Abhishek Das and Ammar Rizvi and Sushree Jagriti Sahoo and Zachary W. Ulissi and C. Lawrence Zitnick},
      year={2025},
      eprint={2506.23971},
      archivePrefix={arXiv},
      primaryClass={cs.LG},
      url={https://arxiv.org/abs/2506.23971}, 
}

@book{Hilfiker2006PolymorphismIndustry,
  author    = {Hilfiker, Rolf (ed.)},
  title     = {Polymorphism in the Pharmaceutical Industry},
  publisher = {Wiley‐VCH},
  address   = {Weinheim},
  year      = {2006},
  note      = {Chaps. 2–4 survey solvent/temperature/pressure screens, mechanochemistry, melt/quench routes, seeding and vapor techniques}
}

@Article{ani1,
author ="Smith, J. S. and Isayev, O. and Roitberg, A. E.",
title  ="ANI-1: an extensible neural network potential with DFT accuracy at force field computational cost",
journal  ="Chem. Sci.",
year  ="2017",
volume  ="8",
issue  ="4",
pages  ="3192-3203",
publisher  ="The Royal Society of Chemistry",
doi  ="10.1039/C6SC05720A",
url  ="http://dx.doi.org/10.1039/C6SC05720A"}

@article{force_balance,
author = {Wang, Lee-Ping and Martinez, Todd J. and Pande, Vijay S.},
title = {Building Force Fields: An Automatic, Systematic, and Reproducible Approach},
journal = {The Journal of Physical Chemistry Letters},
volume = {5},
number = {11},
pages = {1885-1891},
year = {2014},
doi = {10.1021/jz500737m},
note ={PMID: 26273869},
URL = {https://doi.org/10.1021/jz500737m},
eprint = {https://doi.org/10.1021/jz500737m}
}

@article{beran2022interplay,
  title={The interplay of intra-and intermolecular errors in modeling conformational polymorphs},
  author={Beran, Gregory JO and Wright, Sarah E and Greenwell, Chandler and Cruz-Cabeza, Aurora J},
  journal={The Journal of chemical physics},
  volume={156},
  number={10},
  year={2022},
  publisher={AIP Publishing}
}

@article{mcdonagh2019machine,
  title={Machine-learned fragment-based energies for crystal structure prediction},
  author={McDonagh, David and Skylaris, Chris-Kriton and Day, Graeme M},
  journal={Journal of chemical theory and computation},
  volume={15},
  number={4},
  pages={2743--2758},
  year={2019},
  publisher={ACS Publications}
}

@article{beran2010predicting,
  title={Predicting organic crystal lattice energies with chemical accuracy},
  author={Beran, Gregory JO and Nanda, Kaushik},
  journal={The Journal of Physical Chemistry Letters},
  volume={1},
  number={24},
  pages={3480--3487},
  year={2010},
  publisher={ACS Publications}
}

@article{otero2020many,
  title={What is “many-body” dispersion and should I worry about it?},
  author={Otero-de-la-Roza, Alberto and LeBlanc, Luc M and Johnson, Erin R},
  journal={Physical chemistry chemical physics},
  volume={22},
  number={16},
  pages={8266--8276},
  year={2020},
  publisher={Royal Society of Chemistry}
}

@article{giannozzi2020quantum,
  title={Quantum ESPRESSO toward the exascale},
  author={Giannozzi, Paolo and Baseggio, Oscar and Bonf{\`a}, Pietro and Brunato, Davide and Car, Roberto and Carnimeo, Ivan and Cavazzoni, Carlo and De Gironcoli, Stefano and Delugas, Pietro and Ferrari Ruffino, Fabrizio and others},
  journal={The Journal of chemical physics},
  volume={152},
  number={15},
  year={2020},
  publisher={AIP Publishing}
}

@misc{CCDC_CSD_2025,
  author       = {{Cambridge Crystallographic Data Centre}},
  title        = {Cambridge Structural Database: Organic‐Only Space‐Group Frequency Report},
  year         = {2025},
  note         = {Statistics retrieved from the CSD online interface; accessed 19 July 2025},
  url          = {https://www.ccdc.cam.ac.uk/structures}
}

@article{brock1994towards,
  title={Towards a grammar of crystal packing},
  author={Brock, Carolyn Pratt and Dunitz, Jack D},
  journal={Chemistry of materials},
  volume={6},
  number={8},
  pages={1118--1127},
  year={1994},
  publisher={ACS Publications}
}

@article{colwell1994estimating,
  title={Estimating terrestrial biodiversity through extrapolation},
  author={Colwell, Robert K and Coddington, Jonathan A},
  journal={Philosophical Transactions of the Royal Society of London. Series B: Biological Sciences},
  volume={345},
  number={1311},
  pages={101--118},
  year={1994},
  publisher={The Royal Society London}
}

@article{chisholm2005compack,
  title={COMPACK: a program for identifying crystal structure similarity using distances},
  author={Chisholm, James Alexander and Motherwell, Sam},
  journal={Applied Crystallography},
  volume={38},
  number={1},
  pages={228--231},
  year={2005},
  publisher={International Union of Crystallography}
}

@article{li1987monte,
  title={Monte Carlo-minimization approach to the multiple-minima problem in protein folding.},
  author={Li, Zhenqin and Scheraga, Harold A},
  journal={Proceedings of the National Academy of Sciences},
  volume={84},
  number={19},
  pages={6611--6615},
  year={1987}
}

@misc{rdkit,
  author       = {{RDKit Development Team}},
  title        = {{RDKit}: Open‑source cheminformatics},
  year         = {2025},
  howpublished = {\url{https://www.rdkit.org}},
  note         = {[Online; accessed 19 July 2025]}
}

@article{schriber2021cliff,
  title={CLIFF: A component-based, machine-learned, intermolecular force field},
  author={Schriber, Jeffrey B and Nascimento, Daniel R and Koutsoukas, Alexios and Spronk, Steven A and Cheney, Daniel L and Sherrill, C David},
  journal={The Journal of Chemical Physics},
  volume={154},
  number={18},
  year={2021},
  publisher={AIP Publishing}
}

@article{grimme2006semiempirical,
  title={Semiempirical GGA-type density functional constructed with a long-range dispersion correction},
  author={Grimme, Stefan},
  journal={Journal of computational chemistry},
  volume={27},
  number={15},
  pages={1787--1799},
  year={2006},
  publisher={Wiley Online Library}
}

@article{van2016beyond,
  title={Beyond Born--Mayer: Improved models for short-range repulsion in ab initio force fields},
  author={Van Vleet, Mary J and Misquitta, Alston J and Stone, Anthony J and Schmidt, Jordan R},
  journal={Journal of chemical theory and computation},
  volume={12},
  number={8},
  pages={3851--3870},
  year={2016},
  publisher={ACS Publications}
}

@article{ewald1921berechnung,
  title={Die Berechnung optischer und elektrostatischer Gitterpotentiale},
  author={Ewald, Paul P},
  journal={Annalen der physik},
  volume={369},
  number={3},
  pages={253--287},
  year={1921},
  publisher={Wiley Online Library}
}

@article{weenk1977calculation,
  title={Calculation of electrostatic fields in ionic crystals based upon the Ewald method},
  author={Weenk, JW and Harwig, HA},
  journal={Journal of Physics and Chemistry of Solids},
  volume={38},
  number={9},
  pages={1047--1054},
  year={1977},
  publisher={Elsevier}
}

@article{nelson2024convergence,
  title={Convergence of the many-body expansion with respect to distance cutoffs in crystals of polar molecules: Acetic acid, formamide, and imidazole},
  author={Nelson, Philip M and Sherrill, C David},
  journal={The Journal of Chemical Physics},
  volume={161},
  number={21},
  year={2024},
  publisher={AIP Publishing}
}

@article{sargent2023benchmarking,
  title={Benchmarking two-body contributions to crystal lattice energies and a range-dependent assessment of approximate methods},
  author={Sargent, Caroline T and Metcalf, Derek P and Glick, Zachary L and Borca, Carlos H and Sherrill, C David},
  journal={The Journal of Chemical Physics},
  volume={158},
  number={5},
  year={2023},
  publisher={AIP Publishing}
}

@article{liang2023can,
  title={Can spin-component scaled MP2 achieve kJ/mol accuracy for cohesive energies of molecular crystals?},
  author={Liang, Yu Hsuan and Ye, Hong-Zhou and Berkelbach, Timothy C},
  journal={The Journal of Physical Chemistry Letters},
  volume={14},
  number={46},
  pages={10435--10441},
  year={2023},
  publisher={ACS Publications}
}

@article{mardirossian2016omegab97m,
  title={$\omega$B97M-V: A combinatorially optimized, range-separated hybrid, meta-GGA density functional with VV10 nonlocal correlation},
  author={Mardirossian, Narbe and Head-Gordon, Martin},
  journal={The Journal of chemical physics},
  volume={144},
  number={21},
  year={2016},
  publisher={AIP Publishing}
}

@article{jumper2021highly,
  title={Highly accurate protein structure prediction with AlphaFold},
  author={Jumper, John and Evans, Richard and Pritzel, Alexander and Green, Tim and Figurnov, Michael and Ronneberger, Olaf and Tunyasuvunakool, Kathryn and Bates, Russ and {\v{Z}}{\'\i}dek, Augustin and Potapenko, Anna and others},
  journal={nature},
  volume={596},
  number={7873},
  pages={583--589},
  year={2021},
  publisher={Nature Publishing Group UK London}
}

@article{tang1986new,
  title={New combining rules for well parameters and shapes of the van der Waals potential of mixed rare gas systems},
  author={Tang, KT and Toennies, J Peter},
  journal={Zeitschrift f{\"u}r Physik D Atoms, Molecules and Clusters},
  volume={1},
  number={1},
  pages={91--101},
  year={1986},
  publisher={Springer}
}

@article{ponder2002,
author = {Ren, Pengyu and Ponder, Jay W.},
title = {Consistent treatment of inter- and intramolecular polarization in molecular mechanics calculations},
journal = {Journal of Computational Chemistry},
volume = {23},
number = {16},
pages = {1497-1506},
doi = {https://doi.org/10.1002/jcc.10127},
url = {https://onlinelibrary.wiley.com/doi/abs/10.1002/jcc.10127},
eprint = {https://onlinelibrary.wiley.com/doi/pdf/10.1002/jcc.10127},
year = {2002}
}

@article{simmonett2015efficient,
  title={Efficient treatment of induced dipoles},
  author={Simmonett, Andrew C and Pickard, Frank C and Shao, Yihan and Cheatham, Thomas E and Brooks, Bernard R},
  journal={The Journal of chemical physics},
  volume={143},
  number={7},
  year={2015},
  publisher={AIP Publishing}
}

@article{beran2022many,
  title={How many more polymorphs of ROY remain undiscovered},
  author={Beran, Gregory JO and Sugden, Isaac J and Greenwell, Chandler and Bowskill, David H and Pantelides, Constantinos C and Adjiman, Claire S},
  journal={Chemical science},
  volume={13},
  number={5},
  pages={1288--1297},
  year={2022},
  publisher={Royal Society of Chemistry}
}

@article{taylor2025predictive,
  title={Predictive crystallography at scale: mapping, validating, and learning from 1000 crystal energy landscapes},
  author={Taylor, Christopher R and Butler, Patrick WV and Day, Graeme M},
  journal={Faraday Discussions},
  volume={256},
  pages={434--458},
  year={2025},
  publisher={Royal Society of Chemistry}
}

\end{document}